\documentclass[journal]{IEEEtran}

\usepackage{blindtext}

\usepackage{caption}
\usepackage{subcaption}
\usepackage{wrapfig}
\ifCLASSINFOpdf
  \usepackage[pdftex]{graphicx}
\else
\fi
\usepackage{amsmath}
\usepackage{bm}
\usepackage{algorithm,algcompatible}
\usepackage{algpseudocode} % for algorithm

\algnewcommand\Input{\item[\textbf{Input:}]}
\algnewcommand\Output{\item[\textbf{Output:}]}

\usepackage{multirow}
\usepackage{multicol}
\usepackage{xcolor}

\usepackage[hyphens]{url}
\usepackage{hyperref}
\hypersetup{breaklinks=true, urlcolor=blue}
\usepackage{url}
\usepackage{cite}

\usepackage{amsmath}

\usepackage{siunitx} %loaded for writing unit character
\usepackage{nomencl}
\makenomenclature
\begin{document}
\title{Generating Synthetic Systems of Interdependent Critical Infrastructure Networks}
 
\author{Yu~Wang,
        Jin-Zhu~Yu,
        Hiba Baroud % <-this % stops a space
\thanks{The authors are with the Department of Civil and Environmental Engineering, Vanderbilt University, Nashville, TN 37212, USA (email: yu.wang.1@vanderbilt.edu; yujinzhu88@gmail.com; hiba.baroud@vanderbilt.edu)}% <-this % stops a space
% <-this % stops a space
}

\maketitle
\begin{abstract}
The lack of data on critical infrastructure systems has hindered the research progress in modeling and optimizing the system performance. This work develops a method for generating Synthetic Interdependent Critical Infrastructure Networks (SICIN) using simulation and non-linear optimization techniques. SICIN consists of three components: (i) determining the location of facilities in individual networks via a modified simulated annealing algorithm, (ii) generating interdependent links based on a novel pseudo-tripartite graph algorithm, and (iii) simulating network flow using nonlinear optimization considering the operations of individual networks and their interdependencies. Two existing systems of interdependent infrastructure networks are used to validate the proposed method. The results demonstrate that SICIN outperforms state-of-the-art simulation methods according to multiple topological and flow measures of similarity between the simulated and real networks. %SICIN offers an efficient method for generating realistic artificial interdependent infrastructure networks to facilitate future research on the performance modeling and optimization of infrastructure systems. 
The outcomes of this research include (i) a sample of a simulated system of interdependent networks that can serve as a test case for infrastructure models, and (ii) a generalized algorithm that can be used to generate synthetic interdependent infrastructures given partial data of any real network.
\end{abstract}


% \fi
\IEEEpeerreviewmaketitle

\nomlabelwidth=22mm
\nomenclature[01]{\textit{Symbols}}{}
\nomenclature[02]{$\rm{W, P, G}$}{Water, power and gas networks.}
\nomenclature[03]{$\rm{W \to P}$}{Power network depends on water network.}
\nomenclature[04]{$\rm{P \to W}$}{Water network depends on power network.}
\nomenclature[05]{$\rm{G \to P}$}{Power network depends on gas network.}
\nomenclature[06]{$\rm{P \to G}$}{Gas network depends on power network.}
\nomenclature[07]{$\rm{d, t, s}$}{Demand, transmission, supply facilities.}
\nomenclature[08]{$n_{\rm{d}}, n_{\rm{t}}, n_{\rm{s}}$}{Number of demand, transmission and supply nodes.}
\nomenclature[09]{$SS$}{Similarity score.}
\nomenclature[10]{$\tau^m$}{Topological feature of network $m$.}
\nomenclature[11]{$~$}{}
\nomenclature[12]{\textit{Indices and Sets}}{}
\nomenclature[13]{$\mathcal{M}$}{Set of networks, $\mathcal{M} = \\\{\rm{W, P, G, W\to P, G \to P, P\to W, P\to G}\}$.}
\nomenclature[14]{$\mathcal{T}$}{Set of time periods, $\mathcal{T}=\{1,2,...,\mid \mathcal{T} \mid\}$.}
\nomenclature[15]{$\mathcal{N}_{\rm{d}}^m, \mathcal{N}_{\rm{t}}^m, \mathcal{N}_{\rm{s}}^m$}{Set of demand, transmission, and supply nodes in network $m \in \mathcal{M}$.}
\nomenclature[16]{$\mathcal{N}^m$}{Set of nodes in network $m \in \mathcal{M}$, $\mathcal{N}^m = \mathcal{N}_{\rm{d}}^m \cup \mathcal{N}_{\rm{t}}^m \cup \mathcal{N}_{\rm{s}}^m$.}
\nomenclature[17]{$\mathcal{N}^m_{+}(i), \mathcal{N}^m_{-}(i)$}{Set of neighborhood nodes dominated by node $i$ and dominating node $i$ in network $m$.}
\nomenclature[18]{$\mathcal{A}^m$}{Set of arcs in network $m$, arc $a^m_{ij} (i, j\in \mathcal{N}^m)$ points from node $i$ to $j$.}
\nomenclature[19]{$\mathcal{E}^m(i)$}{Set of closest edges relying on node $i$ in network $m$.}
\nomenclature[20]{$\mathcal{S}$}{Set of discrete area segments in region $S$.}
\nomenclature[21]{$~$}{}
\nomenclature[22]{\textit{Decision Variables}}{}
\nomenclature[23]{$f_k^t$}{Flow on arc $k \in \{\mathcal{A}^{\rm{W}}, \mathcal{A}^{\rm{G}}, \mathcal{A}^{\rm{W} \to \rm{P}}, \mathcal{A}^{\rm{G} \to \rm{P}}\}$ at time $t \in \mathcal{T}$.}
\nomenclature[24]{$l_i^t$}{Power load of node $i \in \mathcal{N}^{\rm{P}}$ at time $t \in \mathcal{T}$.}
\nomenclature[25]{$pr_i^t$}{Pressure of node $i \in \{\mathcal{N}^{\rm{G}}, \mathcal{N}^{\rm{P}}_{\rm{s}}\}$ at time $t \in \mathcal{T}$.}
\nomenclature[26]{$~$}{}
\nomenclature[27]{\textit{Matrices and Vectors}}{}
\nomenclature[28]{$\textbf{A}^m$}{Adjacency matrix of network $m$ where $\textbf{A}_{ij}^m = 1$ denotes an arc $a_{ij}^m \in \mathcal{A}^m$.}
\nomenclature[29]{$\textbf{C}^m$}{Closeness matrix of network $m$ where $\textbf{C}_{ij}^m$ is the distance between nodes $i, j\in \mathcal{N}^m$.}
\nomenclature[30]{$\bm{D}^m$}{Degree sequence of network $m$ where $\bm{D}^m_i$ denotes the degree of node $i \in \mathcal{N}^m$.}
\nomenclature[31]{$\bm{r}^m_i$}{Geographic coordinates of the node $i \in \mathcal{N}^m$.}
\nomenclature[32]{$~$}{}
\nomenclature[33]{\textit{Parameters}}{}
\nomenclature[34]{$\varphi(s)$}{Population of area $s \in \mathcal{S}$.}
\nomenclature[35]{$T_0, T_{\rm{min}}$}{The initial and the minimum temperatures controlling simulated annealing.}
\nomenclature[36]{$c, c'$}{Cost of the solution and its updated value in simulated annealing.}
\nomenclature[37]{$\alpha$}{Cooling ratio in simulated annealing.}
\nomenclature[38]{$\lambda$}{The parameter of the Poisson distribution.}
\nomenclature[39]{$a_u, b_u, c_u$}{Fuel consumption coefficients of generating a single unit of electricity per unit of time.}
\nomenclature[40]{$H$}{Heating value of natural gas.}
\nomenclature[41]{$H_{ij}^{ad, t}$}{Enthalpy change of arc $a_{ij} \in \{\mathcal{A}^{\rm{G}}, \mathcal{A}^{\rm{G} \to \rm{P}}\}$.}
\nomenclature[42]{$Z$}{Compressibility factor of natural gas.}
\nomenclature[43]{$Rs$}{Universal gas constant.}
\nomenclature[44]{$T_i$}{Temperature of node $i \in \{\mathcal{N}^{\rm{G}}, \mathcal{N}^{\rm{G} \to \rm{P}}\}$.}
\nomenclature[45]{$\kappa$}{Isentropic exponent.}
\nomenclature[46]{$\eta_{ij}$}{Adiabatic efficiency of the compressor installed at arc $ij \in \{\mathcal{A}^{\rm{G}}, \mathcal{A}^{\rm{G} \to \rm{P}}\}$.}
\nomenclature[47]{$\rho$}{Density of the water.}
\nomenclature[48]{$g$}{Gravitational constant.}
\nomenclature[49]{$h_i$}{Elevation of node $i \in \{\mathcal{N}^{\rm{W}}, \mathcal{N}^{\rm{W} \to \rm{P}}\}$.}
\nomenclature[50]{$\beta$}{Hazen–Williams coefficient.}
\nomenclature[51]{$d_{ij}$}{Diameter of arc (pipe) $a_{ij} \in \{\mathcal{A}^{\rm{W}}, \mathcal{A}^{\rm{G}}\}$.}
\nomenclature[52]{$p_{ij}^{H,t} (p_{ij}^{L,t})$}{Power used to overcome the elevation difference (friction loss) in transporting water along arc $a_{ij} \in \{\mathcal{A}^{\rm{W}}, \mathcal{A}^{\rm{W} \to \rm{P}}\}$.}
\nomenclature[53]{$p_{ij}^t$}{Power consumption of arc $ij \in \mathcal{A}^m, m\in \mathcal{M}$.}
\nomenclature[54]{$K$}{Conversion ratio of the water cooling effect.}
\nomenclature[55]{$z^{m, t}_i$}{Demand for resources at demand node $i\in \mathcal{N}^m_{\rm{d}}, m\in \{\rm{W}, \rm{P}, \rm{G}\}$ at time $t$.}
\nomenclature[56]{$\delta_1- \delta_5$}{Parameters for defining relationship between gas pressure and flow.}
\nomenclature[57]{$e$}{Efficiency factor.}
\nomenclature[58]{$Pr_s, T_s$}{Absolute pressure (520 R) and temperature in standard conditions (14.73 psia).}
\nomenclature[59]{$\chi$}{Ratio of gas molecular weight to that of air.}
\nomenclature[60]{$\phi$}{Gas compressibility}%, which typically decreases as the pressure increases.}
\nomenclature[61]{$c^m$}{Cost of transporting a single unit of resource in network $m\in\{\rm{W}, \rm{P}, \rm{G}\}$.}
\nomenclature[62]{$x^{m,t}$}{Demand of resource $m \in \{\rm{W}, \rm{P}, \rm{G}\}$ per person at time $t \in \mathcal{T}$.}
\printnomenclature

\section{Introduction}
\subsection{Motivation}\label{subsec: motivation}
\IEEEPARstart{I}{nterdependent} Critical Infrastructure (ICI) networks, such as power grids, water and gas distribution networks, telecommunication networks, and multi-modal transportation networks provide essential services to society. As a result of technology development and greater reliance on the services of critical infrastructure, these systems have become increasingly complex \cite{ouyang2009methodological, rinaldi2001identifying, li2008interdependency, qu2020data}. Infrastructure networks interact with each other through interdependencies which are bidirectional interactions across individual networks that influence the overall system operations. Accurate assessment of the system-level performance of interdependent infrastructure networks is essential to the strategic allocation of resources that maintain and protect these systems as well as accelerate their recovery after disruptions \cite{gomez2019integrating, li2018minimax}. Ideally, the system-level performance would be evaluated using real networks where complete information is available on the network topology, spatial characteristics, and operational parameters, such as capacity and flow within individual networks and across networks through interdependent links \cite{ouyang2012three,wang2020quantifying}. However, data on the topology and flow of critical infrastructure are typically not publicly available due to privacy and security concerns \cite{zhang2016modeling}. Additionally, information on the presence and importance of interdependent links is usually not available due to the decentralized management and lack of coordination across different infrastructure sectors~\cite{gomez2019integrating}. \emph{Ex-post} disaster studies have revealed that interdependencies across infrastructure sectors do not behave as expected \cite{bigger2009consequences}. Recent studies emphasize the need for an accurate representation of interdependencies through a better assessment and characterization of uncertainty \cite{reilly2021sources}. However, researchers have had to rely on hypothetical examples and simulations that simplify this representation due to the lack of real data \cite{zhang2016modeling, ouyang2009methodological, fu2016spatial}. Therefore, we develop an approach to generate synthetic data that provide an accurate characterization of interdependent infrastructures and enable improved modeling of their performance.

\subsection{Background}\label{subsec: background}
Prior studies have advanced (i) model-based simulation algorithms to support infrastructure network modeling and (ii) methods to generate synthetic infrastructure data.

For model-based approaches, synthesizing the topology of infrastructure systems is not the immediate objective. However, some studies have relied on the simulation of certain aspects of interdependent infrastructure networks such as network flow \cite{oikonomou2018optimal, he2016robust}. For instance, enhancing the resilience of power-gas networks is achieved using flow optimization under the worst case scenario, resulting in the deviation of the optimized flow from the real one \cite{li2018minimax}. Other studies optimize the use of assets across power and water networks by simulating the system considering only one way dependencies \cite{zamzam2018optimal}. Most model-based simulation approaches classify facilities into two types: the supply facilities that provide resources and the demand facilities that receive and distribute resources to end-users \cite{zhang2016modeling}. To simulate the network topology, supply nodes are randomly distributed in a prescribed area and demand nodes are added sequentially with random coordinates within that area. Newly added nodes are connected to existing ones via undirected links. The initial flow of the simulated network is computed based on nodal betweenness centrality, i.e., the load on each node is computed as its betweenness centrality and the flow along a certain link equals to the number of paths that pass through that link between every pair of supply and demand nodes \cite{hernandez2013probabilistic}. However, most of these studies have only considered physical interdependencies. For example, physical interdependent links between two facilities are modeled based on the geographical distance and the cascading failure along the interdependent link is evaluated as a conditional failure probability \cite{zhang2016modeling}. Other studies consider interdependency as an absolute relationship where a facility cannot operate when the facility it depends on is not operating \cite{almoghathawi2019resilience}. Extensions of this approach include (i) manually adding edges between disconnected components to ensure network connectivity % and a realistic representation of infrastructure networks 
\cite{ouyang2009methodological}, and (ii) incorporating topological and physical constraints governing individual networks (e.g., the gas pipeline model \cite{svendsen2007connectivity} and the DC optimal power flow model \cite{dobson2001initial}).

Approaches focused on generating synthetic infrastructure data have recently garnered attention with methods relying on partial real infrastructure and socio-economic data. These methods do not necessarily represent a real community but are intended to provide researchers access to rich infrastructure data, allowing them to investigate their performance. For instance, Centerville is a virtual city designed with transportation, water/wastewater, and power systems. The systems are generated such that the demographics (e.g., age, income, and employment sectors) are statistically identical to those of a real city. The node layout is contrived to enable certain situations to be tested and the links across infrastructure systems are set up along road links of transportation networks \cite{ellingwood2016centerville}. Other approaches use partial real data to synthesize the network structure for an individual infrastructure. Examples include the IEEE N-bus power and gas systems \cite{grigg1999ieee} as well as recent work on using water demand, source locations, and the road network to synthesize network topology and component characteristics of water distribution networks \cite{ahmad2020synthetic}. However, the majority of these methods do not consider interdependencies, and some studies have assumed that each network has only one type of facility, i.e., all facilities share the same functionality in the system and are distributed with probability proportional to the population density \cite{fu2016spatial}.  %\cite{oikonomou2018optimal} optimize the energy flexibility of water distribution system by linearizing the formalized nonlinear optimization problem. \cite{he2016robust} optimally coordinate electricity and natural gas systems with uncertainties while the computation of two systems is separated by alternating direction method of multipliers.}

While multiple simulation methods have been developed to facilitate research on infrastructure performance modeling, there are several limitations to their ability to model infrastructure operations and interdependencies. First, model-based simulation approaches have made various attempts to characterize interdependencies, however, the strong assumptions made in these methods result in unrealistic simulated ICI systems. These methods assign the locations of nodes randomly, whereas in reality, geographical features and population distribution often determine facility locations. For example, water storage tanks are commonly built near populated areas to ensure timely and cost-effective water distribution while nuclear power plants must be built away from communities. Second, topology-driven approaches hardly guarantee the connectivity of generated networks, which may violate flow conservation rules and potentially lead to infeasible solutions of the flow (e.g., unmet user demand). Finally, the interdependent links in these approaches are first assumed based on domain knowledge and modeled as an absolute relationship even during disruptions, which fails to capture the uncertainty and statistical features of interdependencies (e.g., even though the gas pipelines survive the disruption, they could still lose functionality if the pressure is too low to transport the gas) \cite{reilly2021sources}. Synthesizing infrastructure data depicts a more realistic picture of ICIs by capturing interdependent link operations under uncertainty (e.g., capacity, flow), especially given that information on flows across ICIs is often lacking.

\subsection{Contributions}\label{subsec: contributions}
We address the aforementioned limitations of state-of-the-art approaches by proposing a new method to generate Synthetic Interdependent Critical Infrastructure Networks (SICIN). The outcome includes (i) a fully characterized system of interdependent power, water, and gas networks that can be used to validate and demonstrate models about infrastructure networks, and (ii) the corresponding algorithm that can either be applied or adapted to generate synthetic infrastructure networks. We compare SICIN with state-of-the-art simulation methods using two systems of ICIs. Below is the list of research contributions.
\begin{enumerate}
    \item We formulate the problem of determining facility locations as an optimal distribution problem and develop a modified simulated annealing algorithm to efficiently solve the proposed problem.
    
    \item We devise a pseudo-tripartite graph algorithm to generate infrastructure networks, which ensures the connectivity of generated networks, guarantees the conformity of the associative degree distribution to real networks, and characterizes the supply-transmission-demand level of infrastructure facilities.
    
    \item We simulate the network flow by solving a nonlinear optimization problem that minimizes operational cost. The constraints are strictly designed by considering both operational routines of individual networks and their interdependencies.
\end{enumerate}

The rest of this paper is organized as follows: Section \ref{sec:ici} introduces the ICI system. SICIN is described in Section \ref{sec: topology} (topology simulation) and Section \ref{sec:interdependency and flow} (flow optimization). The numerical results are presented in Section \ref{sec:numerical_simulation}, followed by the conclusion in Section \ref{sec:conclusion}.

\section{Interdependent Critical Infrastructures} \label{sec:ici}
This section describes the power-gas-water networks considered in this study, including their definition, characteristics, and operation. The systems considered throughout this paper are assumed to operate on the city or country levels.
% \begin{figure}[htbp!]
% \centering
% \includegraphics[width=\linewidth]{Interdependent infrastructure simulation tool systems.pdf}
% \caption{Interdependent System of Water-Pwer-Gas networks} 
% \label{Fig: systemcartoon}
% \end{figure}

\begin{figure}[htbp!]
\centering
\includegraphics[width=\linewidth]{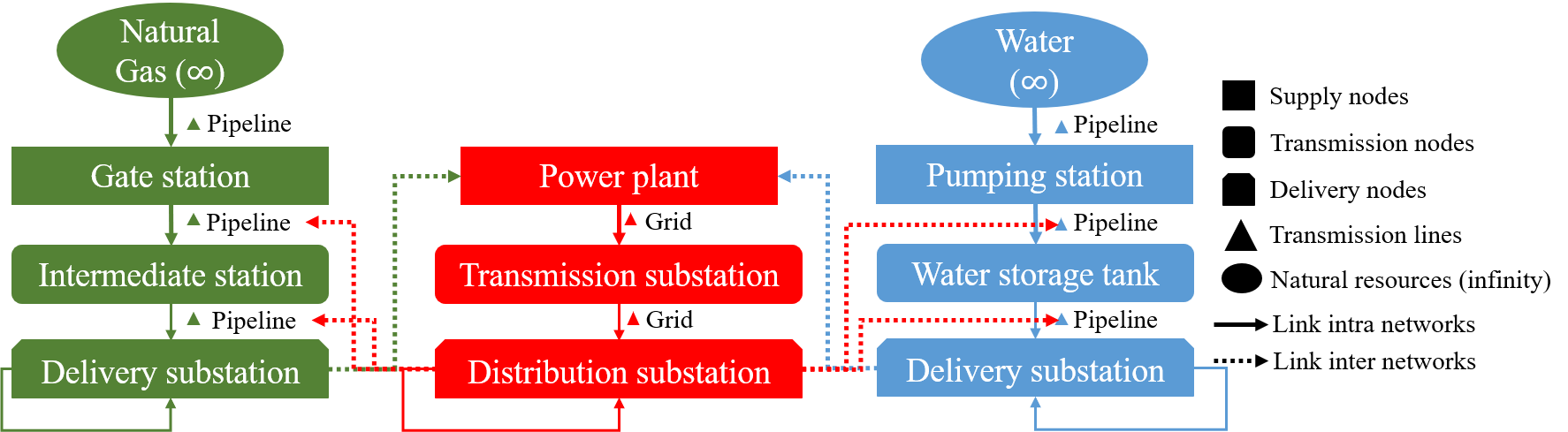}
\caption{Interdependencies across power, water, and gas networks at the county or city level}
\label{Fig: abstract_system}
\end{figure}

In the gas network, the community receives natural gas from external sources (e.g. interstate gas transmission network) through the gate station. Then, the natural gas is transported (driven by the pressure difference generated using electric compressors) through pipelines to users such as households, companies, and the dependent networks (e.g., gas-fired power plants). In power grids, the electricity generated from natural gas is stepped up to a high voltage by transformers and transported to transmission substations where the electric power is then stepped down to a distribution-level voltage. Finally, upon arrival at the service location, the power is stepped down further from the distribution voltage to the service voltage. Besides residential and industrial users, electricity is consumed by other dependent networks. The compressors in gas networks require electricity to generate pressure and transport natural gas. The water network (as a cyber-physical system) requires electricity to distribute water through the use of pumping stations and storage tanks. A portion of water in storage tanks is delivered to power plants for cooling purposes. The operations and interdependencies across the three networks are depicted in Fig.~\ref{Fig: abstract_system}. The interdependencies across the three systems are considered at the county/city level. For many large-scale power and gas systems, the transmission system can span multiple states and countries. The spatial scale of the simulated systems can then be adjusted based on the choice of geographic boundary.
\vspace{-3pt}
\vspace{-3pt}
\\\indent The facilities in each of the three networks are classified into three types: supply, transmission, and demand. Each type is illustrated using different shapes of the nodes as shown in Fig.~\ref{Fig: abstract_system}. Water and gas pipelines and electrical wires are represented by links. The solid lines represent the links within networks and the dashed lines represent links between networks. Links in each network are assumed to only connect nodes of different types, except for the links between demand nodes (i.e., no edge exists between supply nodes or transmission nodes). The natural supply of gas and water is assumed to be unlimited.

The interdependencies are represented as a set of pairwise node-node and node-link dependencies. In node-node dependencies, $\rm{G}_{\rm{d}} \to \rm{P}_{\rm{s}}$ refers to the dependency of power supply nodes on gas demand nodes to generate electricity, and $\rm{W}_{\rm{d}} \to \rm{P}_{\rm{s}}$ refers to the dependency of power supply nodes on water demand nodes for cooling purposes. In node-link dependencies, $\rm{P}_{\rm{d}} \to \rm{G}_{\rm{pipe}}$ refers to the dependency of gas pipelines on electricity from power demand nodes to transport natural gas~\cite{sirvent2017linearized}, and $\rm{P}_{\rm{d}} \to \rm{W}_{\rm{pipe}}$ refers to the dependency of water pipelines on electricity from power demand nodes to transport water flow. The electric power is used to overcome head loss and friction loss. Both water pipelines and gas pipelines consist of intra-network and inter-network links, such as gas pipelines between gas demand nodes and gas pipelines from gas demand nodes to power supply nodes. Within each individual network, flow moves from the supply to demand nodes and customers, whereas in dependent networks such as $\rm{G}_{\rm{d}} \to \rm{P}_{\rm{s}}$, the flow moves from gas demand nodes to power supply nodes which derives from their interconnection (Fig.~\ref{Fig: abstract_system}). A similar interdependence mechanism between gas and power networks is shown in~\cite{ouyang2009methodological} and between power and water networks in~\cite{hernandez2013probabilistic}.
% \begin{table}[htbp!]
% % increase table row spacing, adjust to taste
% \renewcommand{\arraystretch}{1.3}
% \caption{Dependencies in gas-power-water networks}
% \label{table_interdependency}
% \centering
% % Some packages, such as MDW tools, offer better commands for making tables
% % than the plain LaTeX2e tabular which is used here.
% \begin{tabular}{ccc}
% \hline
% \textbf{Name} & \textbf{Head} & \textbf{Tail}\\
% \hline
% Gdemand2Psupply & Gas demand nodes & Power supply nodes\\
% Wdemand2Psupply & Water demand nodes & Power supply nodes\\
% Pdemand2Gpipe & Power demand nodes & Gas pipelines\\
% Pdemand2Wpipe & Power demand nodes & Water pipelines\\
% \hline
% \end{tabular}
% \end{table}
\\\indent Given the operations of individual infrastructure network and their interdependencies, our approach for generating synthetic ICI systems is comprised of three components: (i) determining the location of facilities in individual networks, (ii) generating interdependent links based on a novel pseudo-tripartite graph algorithm, and (iii) simulating network flow using nonlinear optimization. While the method is illustrated using a power-gas-water network, the approach can be adapted to simulate other types of infrastructure networks given the respective topological characteristics and physical laws.

\section{Topology Extraction}\label{sec: topology}
The simulation approach for the topology of interdependent infrastructure networks requires information on the area's geographic boundary, population distribution, and daily consumption to generate (i) node locations and (ii) the adjacency relationship among the nodes.

\subsection{Node Location}\label{subsec: nodelocation}
Transporting resources from supply nodes to demand nodes in infrastructure networks incurs a cost proportional to the distance and amount of flow. To minimize such cost, the node location should be determined such that the overall distance between every user and their nearest node is minimized \cite{gastner2006shape, xie2007geographical, fu2016spatial}. The overall distance is calculated by integrating over the distances between users and their nearest demand nodes. For example, the target network $m$ is located in the area $\mathcal{S}$ and for any subarea $s\in\mathcal{S}$, the population density is $\varphi(s)$, and the geographical coordinate of this subarea is $\bm{r}_{s}$. We calculate the distance between the centroid of this subarea $s$ and its nearest demand node $i$ by $\min_{i\in\mathcal{N}_d^{m}}{||\bm{r}_s - \bm{r}_i||}$, i.e., we find the nearest demand node $i$ from the set of demand nodes $\mathcal{N}_d^m$ in network $m$ and calculate its distance to the centroid. For subarea $s$, the population density is $\varphi(s)$ and its area is $ds$. As such, the total population is $\varphi(s)ds$ and the total distance to access resources in $s$ from the nearest demand node, $i$, is $\min_{i\in\mathcal{N}_d^{m}}{||\bm{r}_s - \bm{r}_i||}\varphi(s)ds$. To get the overall distance of all populations in the area $\mathcal{S}$ to reach their corresponding nearest demand nodes and access resources, we integrate $\min_{i\in\mathcal{N}_d^{m}}{||\bm{r}_s - \bm{r}_i||}\varphi(s)ds$ over all subareas $s\in\mathcal{S}$, which leads to \eqref{eq:distance1}. Similarly, the location of transmission nodes and supply nodes are determined by minimizing the distance from every demand node to the nearest transmission node, and every transmission node to the nearest supply node, \eqref{eq:distance2}-\eqref{eq:distance3}.
\begin{equation}\label{eq:distance1}
f(\bm{r}_{\mathcal{N}_{\rm{d}}^m}) = \int\limits_{\mathcal{S}} {\mathop {\min }\limits_{i \in \mathcal{N}^m_{\rm{d}}} \left\| {\bm{r}_s - \bm{r}_i} \right\|\varphi (s){d}s}, \forall{m \in \mathcal{M}}
\end{equation}
\begin{equation} \label{eq:distance2}
f(\bm{r}_{\mathcal{N}_{\rm{t}}^m}; \bm{r}_{\mathcal{N}_{\rm{d}}^m}) = \sum\limits_{i \in \mathcal{N}^m_{\rm{d}}} {\mathop {\min }\limits_{j \in \mathcal{N}^m_{\rm{t}}} \left\| {\bm{r}_j - \bm{r}_i}\right\|}, \forall{m \in \mathcal{M}}
\end{equation}
\begin{equation}\label{eq:distance3}
f(\bm{r}_{\mathcal{N}_{\rm{s}}^m}; \bm{r}_{\mathcal{N}_{\rm{t}}^m}) = \sum\limits_{i \in \mathcal{N}^m_{\rm{t}}} {\mathop {\min }\limits_{j \in \mathcal{N}^m_{\rm{s}}} \left\| {\bm{r}_j - \bm{r}_i}\right\|}, \forall{m \in \mathcal{M}}
\end{equation}

Constraint \eqref{eq:geographical constriction} imposes geographical restrictions on selecting node locations $\bm{r}_{\mathcal{N}_i^m}$, and $\mathcal{S}$ excludes all infeasible locations such as locations that are already occupied and locations where geological conditions are unsuitable to build infrastructure facilities. Note that specific landscape constraints are not taken into account in order to enable the applicability of our approach in different contexts and areas. Since such constraints vary significantly from one area to another, they can be incorporated by adapting constraint \eqref{eq:geographical constriction} to reflect the area's specific characteristics.
\begin{equation}\label{eq:geographical constriction}
    \bm{r}_{\mathcal{N}_i^m} \subset S, \forall i \in \{\rm{d},\rm{t},\rm{s}\}
\end{equation}

The problem given by \eqref{eq:distance1} to \eqref{eq:geographical constriction} is a p-median problem, which has been proven to be NP-hard~\cite{gastner2006optimal}, i.e., no polynomial-time solution is available. One of the predominant algorithms for solving the p-median problems is metaheuristics, such as tabu search and genetic algorithm (GA)~\cite{reese2006solution}. A meta-heuristic algorithm, \textit{Simulated Annealing} (SA)~\cite{van1987simulated,haddock1992simulation}, is modified and implemented to approximate the global minimum of the overall distance in \eqref{eq:distance1} to \eqref{eq:distance3}. Compared to other metaheuristics, SA has been shown to statistically guarantee an optimal solution for arbitrary problems with a large enough initial temperature and
a proper temperature schedule \cite{geman1984stochastic}. Additionally, SA has been widely used to solve the p-median problem \cite{gastner2005spatial, chiyoshi2000statistical, gastner2006optimal, levanova2004algorithms}. The key feature of the SA algorithm is that it can avoid local minima by allowing hill-climbing moves (moves to a worse solution) probabilistically according to the Metropolis criterion \cite{henderson2003theory}. We modify the original SA by replacing the fixed number of iterations $UB$ at step 3 in Algorithm \ref{alg:anneal_simulation} with an iteration variable $UB(T_0)$ that increases monotonically as the temperature $T_0$ decreases. Furthermore, we reduce the search space as the temperature increases. A smaller number of iterations with larger search space at higher temperature allows the full exploration of the feasible space and thus increases the chance of finding the neighboring area of the optimal solution, i.e., coarse search with higher variance for promising solutions. As the temperature decreases, the number of iterations increases with a smaller search space to promote a fine search, allowing the algorithm to precisely locate the optimal solution in neighboring areas.

%old
% \begin{equation} \label{eq:prob}
%     p=e^{\frac{c-c'}{T}}
% \end{equation}
% In~\eqref{eq:prob}, $c - c'$ represents the energy difference which is the change in the cost function between two sequential iterations, and $T$ denotes the current temperature. Typically, the temperature starts at one and decreases gradually as the optimization proceeds. Since determining locations of transmission and supply nodes is the same as demand nodes, we only present the simulated annealing algorithm for demand nodes here. The original simulated annealing algorithm is modified by replacing the fixed number of iterations $UB$ at step 3 in Algorithm~\ref{alg:anneal_simulation} with an iteration number $UB(T)$ as a function of the temperature $T$, which monotonically increases as the temperature decreases. A smaller number of iterations at high temperature allows full exploration of the feasible space, i.e., coarse search for promising solutions. As temperature decreases, the number of iterations increases to promote a fine search, allowing the algorithm to find the global optimal assignment of node locations around the promising solutions.
\begin{algorithm}[htbp!]
    \caption{Modified simulated annealing (MSA) for optimizing the location of demand nodes}
    \label{alg:anneal_simulation}
    \begin{algorithmic}[1]
   \Input{area $\mathcal{S}$, population distribution $\varphi(\mathcal{S})$, the number of demand nodes $n_{\rm{d}}$, the initial and the minimum temperature $T_0, T_{\rm{min}}$, the cooling ratio $\alpha$, and the upper bound of the iteration number at each temperature $UB(T_0)$}
   \Output{location of demand nodes $\bm{r}_{\mathcal{N}^m_{\rm{d}}}$}
   \State Construct the initial feasible solution by randomly assigning node coordinates within the area $S$ and calculate the cost $c$ by~\eqref{eq:distance1}
    \While{$T_0 > T_{\rm{min}}$}
        \For{$i\gets 1$ to $UB(T_0)$}
            \State Generate a new feasible solution by changing the location of a randomly selected demand node; the search space reduces as the temperature increases
            \State Calculate the new cost $c'$ by \eqref{eq:distance1} and the acceptance probability $p=\exp(\frac{c-c'}{T})$
            \State Generate a random number $j\sim U(0,1)$
            \If{$p >= j$}
                \State Update the solution and the cost $c\gets c'$
            \EndIf
        \EndFor
        \State $T_0 \gets T_0\times \alpha$
    \EndWhile
    \State \textbf{return} $\bm{r}_{\mathcal{N}_{\rm{d}}^m}$
\end{algorithmic}
\end{algorithm}

\vspace{-8pt}

\subsection{Adjacency relationship}\label{subsec:adjmatrix}
Once the node location is determined, the adjacency relationship among nodes is required to simulate the network topology. Most existing approaches generate links among nodes based on the spatial proximity and the node degree \cite{zhang2016modeling, almoghathawi2019resilience, ouyang2009methodological, casey2005self}. However, this method fails to guarantee that (i) simulated networks are connected, and (ii) their topology features are similar to real infrastructure networks. For instance, the degree distribution of the simulated network often does not match the typical Poisson distribution of infrastructure networks \cite{giustolisi2017network}. Therefore, we develop a pseudo-tripartite graph algorithm to simulate ICIs with the prescribed degree distribution obtained from real infrastructure networks. Each individual infrastructure network is treated as a pseudo-tripartite graph wherein the three partitions correspond to supply-transmission, transmission-demand, and demand-demand. The adoption of the tripartite graph structure derives from the fact that (i) nodes in most infrastructure networks are categorized into supply, transmission, and demand nodes depending on the function of the facilities represented by the nodes, and (ii) edges represent pipelines or power lines that transport resources between nodes of different types. Similar types of infrastructure display particular properties, such as degree distribution \cite{guimera2004modeling}. Thus, node degree distribution is extracted from real and similar types of networks to produce realistic topological features of simulated networks. The procedure for generating pseudo-tripartite graph with prescribed nodal degree distribution is summarized in Algorithm \ref{alg:tripartite_graph}.

\indent In both the real infrastructure network $m'$ and the network to be simulated $m$, the numbers of supply nodes are $n_{\rm{s}}^{m'}, n_{\rm{s}}^m$, the numbers of transmission nodes are $n_{\rm{t}}^{m'}, n_{\rm{t}}^m$, and the numbers of demand nodes are $n_{\rm{d}}^{m'}, n_{\rm{d}}^{m}$. Steps 1-2 fit a Poisson distribution to the degree sequence $\bm{D}^{m'}$ of the network $m'$. Step 3 samples a degree sequence $\bm{D}^m$ from the fitted Poisson distribution. Steps 4-15 generate the network $m$ of tripartite structure using the sampled degree sequence $\bm{D}^m$, where $\{\mathcal{N}_{\rm{s}}, \mathcal{N}_{\rm{t}}\}$ and $\{\mathcal{N}_{\rm{t}}, \mathcal{N}_{\rm{d}}\}$ correspond to supply-transmission and transmission-demand partitions. Steps 9-14 recheck the connectivity of the network, and once an isolated node is found in $\mathcal{N}_2$, an edge is added between that node and its nearest node in $\mathcal{N}_1$. The name 'pseudo' derives from the fact that when sets $\{\mathcal{N}_1, \mathcal{N}_2\} = \{\mathcal{N}_{\rm{d}}, \mathcal{N}_{\rm{d}}\}$, steps 5-8 add edges between demand nodes and the resulting graph is not tripartite. Adding edges between demand nodes is a reasonable approach since some resources are transported to demand nodes via other demand nodes instead of directly from supply or transmission nodes. The generated network $m$ is acyclic and its degree distribution is approximately the same as the real infrastructure network $m'$. It should be noted that the degree sequence used to fit the Poisson distribution $\bm{D}^{m'}$ comes from the infrastructure network $m'$, which is required to be of the same type of the target network $m$. Given that the degree distribution is shared by similar infrastructure sectors, data from any given network of the same type of infrastructure, $m'$, can be used to fit the parameters of the Poisson distribution. The time complexity for this algorithm is near-linear and thus well-accepted. Step 1 takes $O(|\mathcal{N}^{m}| + |\mathcal{A}^{m}|)$ to calculate the degree sequence of network $m$. Distribution fitting in Step 2 is fast using existing functions in any computing platform. Step 3 takes $O(1)$ to sample from the continuous Poisson distribution. Steps 4-16 perform nearest neighbor search for every node, which takes $O(|\mathcal{N}^{m}|^2)$ if using the naive brute force search and can be further reduced to $O(|\mathcal{N}^{m}|\log(|\mathcal{N}^{m}|))$ using KD-tree \cite{wald2006building}. The total time complexity is then $O\left(|\mathcal{N}^{m}|\log(|\mathcal{N}^{m}|) + |\mathcal{N}^{m}| + |\mathcal{A}^{m}|\right)$.

\begin{algorithm}[htbp!]
    \caption{Generating the pseudo-tripartite graph with prescribed nodal degree distribution}\label{alg:tripartite_graph}
    \begin{algorithmic}[1]
  \Input{the adjacency matrix $\textbf{A}^{m'}$ of the real infrastructure network $m'$, the number of supply nodes $n_{\rm{s}}$, transmission nodes $n_{\rm{t}}$, and demand nodes $n_{\rm{d}}$ of the simulated infrastructure network (the pseudo-tripartite graph) $m$, the sets of supply $\mathcal{N}^m_{\rm{s}}$, transmission nodes $\mathcal{N}^m_{\rm{t}}$, and demand nodes $\mathcal{N}^m_{\rm{d}}$, the parameter of the Poisson distribution to be fitted $\lambda$, the closeness matrix $\textbf{C}^m$}
  \Output{the adjacency matrix $\textbf{A}^m$ of size $|\mathcal{N}^m| \times |\mathcal{N}^m|$}
  \State Obtain the degree sequence $\bm{D}^{m'}$ of network $m'$ by $\bm{D}_i^{m'}=\sum^{|\mathcal{N}^{m'}|}_{j=1}{\textbf{A}^{m'}_{ij}}, \forall i \in \mathcal{N}^{m'}$
  \State Fit a Poisson distribution with parameter $\lambda$ to the degree sequence data $\bm{D}^{m'}$
  \State Sample a degree sequence $\bm{D}^m$ of size $\mathcal{N}^m$ from the fitted distribution as the degree sequence of network $m$
  \For{sets $\{\mathcal{N}_1, \mathcal{N}_2\} \in \{\{\mathcal{N}_{\rm{s}}, \mathcal{N}_{\rm{t}}\},\{\mathcal{N}_{\rm{t}}, \mathcal{N}_{\rm{d}}\}, \{\mathcal{N}_{\rm{d}}, \mathcal{N}_{\rm{d}}\}\}$}
    \For{node $i \in \mathcal{N}_1$}
        \State Find the nearest $\min\{\bm{D}^m_i, \left|{\mathcal{N}_2}\right|\}$ nodes to node $i$
        \State and connect them with node $i$ 
        \State Update the adjacency matrix by  $\textbf{A}^m_{ij}\leftarrow 1$
    \EndFor
    \For{node $i \in \mathcal{N}_2$}
        \If{$\mathcal{N}_1$ contains no nodes pointing to node $i$}
            \State Find the nearest node $j$ to node $i$ from $\mathcal{N}_1$ and
            \State set $\textbf{A}^m_{ji} \leftarrow 1$
        \EndIf
    \EndFor
    \EndFor
    \State \textbf{return} $\textbf{A}^m$
\end{algorithmic}
\end{algorithm}

\section{Modeling Interdependency and Flow} \label{sec:interdependency and flow}
Current simulation methods and infrastructure models assume that interdependencies are fixed and binary, whereas in reality, these links are uncertain and dynamic, especially during disasters \cite{reilly2021sources,yu2020modeling}. We propose to generate interdependent links along with their corresponding capacity given the network topology and flow initialization. The approach is formulated as an optimization problem that minimizes the network operational cost subject to physical constraints of the three networks and their interdependencies.

\subsection{Interdependent Links}
The physical links for dependencies $\rm{G}_{\rm{d}} \to \rm{P}_{\rm{s}}, \rm{W}_{\rm{d}} \to \rm{P}_{\rm{s}}, {\rm{P}_{\rm{d}} \to \rm{G}_{\rm{pipe}}}$, and $\rm{P}_{\rm{d}} \to \rm{W}_{\rm{pipe}}$ illustrated in Section~\ref{sec:ici} are added based on geographic proximity (Euclidean distance)~\cite{zhang2016modeling, almoghathawi2019resilience, mooney2018facility}. For instance, to build $\rm{G}_{\rm{d}} \to \rm{P}_{\rm{s}}$ dependency, gas pipelines are added between power supply nodes and the nearest gas demand nodes. To build node-to-link dependencies, the distance between the node and the link is approximated by the distance from the middle point of the link. 

\subsection{Flow Constraints for Interdependent Links} \label{subsec: interdependency characterization}
Interdependencies between gas and power networks are the result of (i) the dependency of power supply nodes on the fuel provided by gas demand nodes to generate electricity, (i.e., $\rm{G}_{\rm{d}} \to \rm{P}_{\rm{s}}$), and (ii) reliance of compressor machines in gas pipelines on electricity from power demand nodes to increase pressure and transport gas over long distances (i.e., $\rm{P}_{\rm{d}} \to \rm{G}_{\rm{pipe}}$) ~\cite{sirvent2017linearized}.
\\\indent For every power supply node $i \in \mathcal{N}_{\rm{s}}^{\rm{P}}$, the total amount of natural gas transported at time $t$ from its dependent gas demand nodes $j \in \mathcal{N}_ - ^{\rm{G} \to \rm{P}}(i)$ is burned to generate $H\sum\limits_{j \in \mathcal{N}_ - ^{\rm{G} \to \rm{P}}(i)} {f_{ji}^{t}}$ energy with the heating value of natural gas $H$. Additionally, a certain amount of energy is further converted to generate $l_i^t$ units of electricity in the quadratic function with fuel consumption coefficients $a_u, b_u, c_u$ according to ~\eqref{eq:interg2p}, ~\cite{sirvent2017linearized}.
\begin{equation}\label{eq:interg2p}
    \frac{{{a_u} + {b_u}{l_i^{t}} + {c_u}(l_i^{t})^2}}{H} = \sum\limits_{j \in \mathcal{N}_ - ^{\rm{G} \to \rm{P}}(i)} {f_{ji}^{t},\forall i}  \in \mathcal{N}_{\rm{s}}^{\rm{P}}
\end{equation}
\indent For every gas pipeline $(j, k) \in (\mathcal{A}^{\rm{G}}\cup \mathcal{A}^{\rm{G} \to \rm{P}})$, the power consumption for the compressor to increase pressure and transport gas is given by~\eqref{eq:interp2g1}-\eqref{eq:interp2g2}, ~\cite{sirvent2017linearized}. The change in adiabatic enthalpy, $H^{ad, t}_{jk}$, is calculated using \eqref{eq:interp2g1} based on the pressure at two endpoints of the gas pipeline, $pr_k^t$ and $pr_j^t$, which is then used to calculate the power consumption of the pipeline, $p^t_{jk}$, ~\eqref{eq:interp2g2}. The power consumption $p_{jk}^t$ is directly loaded onto its corresponding power demand node $i \in \mathcal{N}^{\rm{P}}_{\rm{d}}$ and is used to determine the status of power balance in power networks, described in~\ref{subsec:DCpowerflow}.
\begin{align}\label{eq:interp2g1}
H_{jk}^{ad,t} &= \frac{{ZRs{T_j}}}{{(\kappa  - 1)/\kappa }}({(\frac{{pr_j^t}}{{pr_k^t}})^{(\kappa  - 1)/\kappa }} - 1), \nonumber \\&\forall (j,k) \in (\mathcal{A}^{\rm{G}}\cup \mathcal{A}^{\rm{G} \to \rm{P}})
\end{align}
\begin{equation}\label{eq:interp2g2}
p_{jk}^t = \frac{{f_{jk}^tH_{jk}^{ad,t}}}{{33000{\eta _{jk}}}}, \forall (j,k) \in (\mathcal{A}^{\rm{G}}\cup \mathcal{A}^{\rm{G} \to \rm{P}})
\end{equation}
\indent Pumping stations depend on the electricity provided by the power demand nodes to extract water from nearby rivers and then transport the water through water pipelines to storage tanks and end-users. For every water pipeline $(j, k) \in (\mathcal{A}^{\rm{W}}\cup \mathcal{A}^{\rm{W} \to \rm{P}})$, two types of power loss could occur when transporting the water, (i) the headloss $p_{jk}^{H,t}$, and (ii) the friction loss $p_{jk}^{L,t}$. The headloss arises from overcoming the gravitational energy of transporting water from a node at elevation $h_j$ to a node at elevation $h_k$, ~\eqref{eq:interWP1}. The friction loss results from the energy consumption due to pipeline roughness, and the Hazen-Williams equation is employed,~\eqref{eq:interWP2}~\cite{shuang2014node}. Similar to the dependency of gas pipelines on power demand nodes, the power loss experienced in water pipelines is made up using the corresponding nearest power demand node $i \in \mathcal{N}^{\rm{P}}_{\rm{d}}$.
\begin{equation}\label{eq:interWP1}
    p_{jk}^{H, t} = \rho gf_{jk}^t({h_k} - {h_j}),\forall (j,k) \in ({{\cal A}^{\rm{W}}} \cup {{\cal A}^{\rm{W} \to \rm{P}}})
\end{equation}
\begin{equation}\label{eq:interWP2}
    p_{jk}^{L,t} = 10.654{(\frac{{f_{jk}^t}}{\beta })^{1.852}}\frac{{\textbf{C}_{jk}^m}}{{{d_{jk}}}},\forall (j,k) \in ({{\cal A}^{\rm{W}}} \cup {{\cal A}^{\rm{W} \to \rm{P}}})
\end{equation}
\begin{equation}\label{eq:interWP3}
    p_{jk}^t = p_{jk}^{H,t} + p_{jk}^{L,t},\forall (j,k) \in ({{\cal A}^{\rm{W}}} \cup {{\cal A}^{\rm{W} \to \rm{P}}})
\end{equation}
\indent In power plants, natural gas is burned to heat water and produce steam that drives the turbines to generate electricity. Then, large volumes of water are withdrawn from nearby rivers, lakes, and oceans to cool the steam back into liquid water. Assuming the conversion ratio is $K$ (i.e., generating each unit of electricity requires $K$ units of water), this dependency can be quantified using~\eqref{eq:interWP4}.
% On average, producing 1 kilowatt-hour of electricity consumes 95L water \cite{jones2008much}. With this conversion ratio, we have:
\begin{equation}\label{eq:interWP4}
    \sum\limits_{i \in \mathcal{N}_ - ^{{\rm{W} \to \rm{P}}}(j)} {f_{ij}^t} = K l_j^t,\forall j \in \mathcal{N}_{\rm{s}}^{\rm{P}}
\end{equation}

\subsection{Flow Constraints for Individual Networks} \label{subsec: individual network}
This subsection describes the constraints for the water, gas, and power networks, respectively.
\subsubsection{Hydraulics Modeling of Water Flow}
\begin{equation}\label{eq:wflowconservetran}
    \sum\limits_{j \in \mathcal{N}_ - ^{\rm{W}}(i)} {f_{ji}^t}  = \sum\limits_{j \in \mathcal{N}_ + ^{\rm{W}}(i)} {f_{ij}^t}, \forall i \in \mathcal{N}_{\rm{t}}^{\rm{W}}
\end{equation}
\begin{equation}\label{eq:wflowconservedemand}
    \sum\limits_{j \in \mathcal{N}_ - ^{\rm{W}}(i)} {f_{ji}^t}  = \sum\limits_{j \in \mathcal{N}_ + ^{\rm{W}}(i)} {f_{ij}^t}  + \sum\limits_{j \in \mathcal{N}_ + ^{{\rm{W} \to \rm{P}}}(i)} {f_{ij}^t}  + z_i^{{\rm{W}}, t},\forall i \in {\mathcal{N}}_{\rm{d}}^{\rm{W}}
\end{equation}
\indent Constraints~\eqref{eq:wflowconservetran}-\eqref{eq:wflowconservedemand} ensure the flow conservation at water transmission nodes $\mathcal{N}^{\rm{W}}_{\rm{t}}$ and water demand nodes $\mathcal{N}^{\rm{W}}_{\rm{d}}$. Specifically for water demand nodes, residents' water demand $z_i^{{\rm{W}}, t}$ is incorporated as the extra sink in the flow conservation.
\subsubsection{Pipeline Modeling of Gas Flow}
\begin{equation}\label{eq:gflowconservetran}
    \sum\limits_{j \in \mathcal{N}_ - ^{\rm{G}}(i)} {f_{ji}^t}  = \sum\limits_{j \in \mathcal{N}_ + ^{\rm{G}}(i)} {f_{ij}^t} , \forall i \in \mathcal{N}_{\rm{t}}^{\rm{G}}
\end{equation}
\begin{equation}\label{eq:gflowconservedemand}
    \sum\limits_{j \in \mathcal{N}_ - ^{\rm{G}}(i)} {f_{ji}^t}  = \sum\limits_{j \in \mathcal{N}_ + ^{\rm{G}}(i)} {f_{ij}^t}  + \sum\limits_{j \in \mathcal{N}_ + ^{{\rm{G} \to \rm{P}}}(i)} {f_{ij}^t}  + z_i^{{\rm{G}}, t},\forall i \in {\mathcal{N}}_{\rm{d}}^{\rm{G}}
\end{equation}
\begin{align}\label{eq:gaspressureflow}
    f_{ij}^t = {\delta _1}e{({d_{ij}})^{{\delta _2}}}{(\frac{{{T_s}}}{{P{r_s}}})^{{\delta _3}}}{(\frac{{{{(pr_i^t)}^2} - {{(pr_j^t)}^2}}}{{{{\chi}^{{\delta _4}}}{\bf{C}}_{ij}^m{T_i}\phi }})^{{\delta _5}}}, \nonumber&\\\forall (i,j) \in ({{\cal A}^{\rm{G}} \cup {\cal A}^{\rm{G} \to \rm{P}}})
\end{align}
\indent Similarly to constraints~\eqref{eq:wflowconservetran}-\eqref{eq:wflowconservedemand}, constraints~\eqref{eq:gflowconservetran}-\eqref{eq:gflowconservedemand} ensure the flow conservation at transmission and demand nodes in gas networks. Constraint~\eqref{eq:gaspressureflow} is the Weymouth equation restricting the relationship between the pressure drop and the amount of gas flow along pipelines~\cite{crane1988flow, alabdulwahab2015stochastic}.
\subsubsection{DC Power Flow Model}\label{subsec:DCpowerflow}
\begin{align} \label{eq:power cal}
    {l_i^t} &= z_i^{{\rm{P}},t} + \sum\limits_{(j, k) \in \mathcal{E}^{\rm{W}}(i)} {{p_{jk}^t}}  + \sum\limits_{(j, k) \in \mathcal{E}^{\rm{G}}(i)} {{p_{jk}^t}} \nonumber\\&+ \sum\limits_{(j, k) \in \mathcal{E}^{\rm{W} \to \rm{P}}(i)} {{p_{jk}^t}} + \sum\limits_{(j, k) \in \mathcal{E}^{\rm{G} \to \rm{P}}(i)} {{p_{jk}^t}} ,\forall i \in \mathcal{N}_{\rm{d}}^{\rm{P}}
\end{align}
\begin{equation}\label{eq:edge1}
    \bigcup_{i \in \mathcal{N}_{\rm{d}}^{\rm{P}}}{\mathcal{\mathcal{E}}^{\rm{W}}(i)} = \mathcal{A}^{\rm{W}}
\end{equation}
\begin{equation}\label{eq:edge2}
    \bigcup_{i \in \mathcal{N}_{\rm{d}}^{\rm{P}}}{\mathcal{\mathcal{E}}^{\rm{G}}(i)} = \mathcal{A}^{\rm{G}}
\end{equation}
\begin{equation}\label{eq:edge3}
    \bigcup_{i \in \mathcal{N}_{\rm{d}}^{\rm{P}}}{\mathcal{\mathcal{E}}^{\rm{P} \to \rm{W}}(i)} = \mathcal{A}^{\rm{P} \to \rm{W}}
\end{equation}
\begin{equation}\label{eq:edge4}
    \bigcup_{i \in \mathcal{N}_{\rm{d}}^{\rm{P}}}{\mathcal{E}^{\rm{P} \to \rm{G}}(i)} = \mathcal{A}^{\rm{P} \to \rm{G}}
\end{equation}
\begin{equation}\label{eq:powerbalance}
    \sum\limits_{i \in \mathcal{N}_{\rm{s}}^{\rm{P}}} {{l_i^t}}  = \sum\limits_{i \in \mathcal{N}_{\rm{d}}^{\rm{P}}} {{l_i^t}}
\end{equation}
% \begin{equation}\label{eq:DCpowerflow}
%     \boldsymbol{p} = \mathbf{B}\boldsymbol{\theta}
% \end{equation}
% \begin{equation}\label{eq:Bcalculate}
%     {B_{ij}} = \left\{ {\begin{array}{*{20}{c}}
% {\sum\limits_{k \in N_ + ^P(i)} {{b_{ik}},i = j} }\\
% { - {b_{ij}},~~~~~~i \ne j}
% \end{array}} \right
% \end{equation}
% \begin{equation}\label{eq:powerflow}
%     f_{ij} = b_{ij}({\theta _i} - {\theta _j}),\forall (i,j) \in ({{\cal E}^P}\cup {{{\cal E}^{{I^{P \to W}}}}} \cup {{{\cal E}^{{I^{P \to GS}}}}} )
% \end{equation}
\indent For each power demand node $i \in \mathcal{N}^{\rm{P}}_{\rm{d}}$, \eqref{eq:power cal} considers all sources of power consumption to calculate the total power load, including power for transporting natural gas~\eqref{eq:interp2g1}-\eqref{eq:interp2g2}, power for transporting water~\eqref{eq:interWP1}-\eqref{eq:interWP3}, and power to serve residents' demand $z_i^{{\rm{P}}, t}$. Constraints \eqref{eq:edge1}-\eqref{eq:edge4} ensure that every pipeline in the water and natural gas networks is served by at least one power demand node. Similar to the conservation constraints in water and gas networks, \eqref{eq:powerbalance} ensures the overall power demand is met by the overall power supply. To obtain the power load, the power flow along power lines, $f_{ij}^t$, is usually evaluated using the voltage angle $\theta$ calculated by multiplying $l$ by the inverse of the susceptance matrix. However, this calculation is omitted since the power flow is solved independently of the flow optimization. More details about the DC power flow model can be found in~\cite{dobson2001initial}.

\subsection{System-Level Optimization} \label{subsec: system-level}
The power, gas, and water networks are expected to operate according to each network's physical laws and constraints of interdependencies. To minimize the total operational cost~\eqref{eq:obj} under the interdependency and individual network constraints defined previously, the system-level optimization problem is formulated as follows:
\begin{align}
	\min \quad & {c^{\rm{W}}}(\sum\limits_{(i,j) \in {\mathcal{A}^{\rm{W}}}} {{f_{ij}^t}{\textbf{C}^{\rm{W}}_{ij}}} + \sum\limits_{(i,j) \in {\mathcal{A}^{\rm{W} \to \rm{P}}}} {{f_{ij}^t}{\textbf{C}^{\rm{W} \to \rm{P}}_{ij}}})  \nonumber+ \\&{c^{\rm{G}}}(\sum\limits_{(i,j) \in \mathcal{A}^{\rm{G}}} {{f_{ij}^t}{\textbf{C}_{ij}^{\rm{G}}}} +\sum\limits_{(i,j) \in {{\mathcal{A}^{\rm{G} \to \rm{P}}}}} {{f_{ij}^t}{\textbf{C}_{ij}^{\rm{G} \to \rm{P}}}})  + {c^{\rm{P}}}\sum\limits_{i \in \mathcal{N}_{\rm{s}}^{\rm{P}}} {{p_i^t}} \label{eq:obj}\\
	\text{s.t.}  \quad & \text{P-G interdependency constraints}~(\ref{eq:interg2p})-(\ref{eq:interp2g2}) \nonumber\\
	& \text{W-P interdependency constraints}~(\ref{eq:interWP1})-(\ref{eq:interWP4})\nonumber\\
	& \text{water constraints}~(\ref{eq:wflowconservetran})-(\ref{eq:wflowconservedemand})\nonumber\\
	& \text{natural gas constraints}~(\ref{eq:gflowconservetran})-(\ref{eq:gaspressureflow})\nonumber\\
	& \text{power constraints}~(\ref{eq:power cal})-(\ref{eq:powerbalance})\nonumber
\end{align}

Since constraints (5)-(6), (9), and (16) are not affine, this nonlinear optimization problem is nonconvex, which can be solved using the relaxation method \cite{boyd1997semidefinite} and the multi-start randomized method (MSRM) \cite{jain1989global}. Given that solving for the suboptimaltiy of our problem is fast, we can obtain numerous suboptimal solutions from different initialization points and approximate the global optimal by the minimum among the local optimal solutions. As the number of suboptimal solutions from different initialization points increases, the approximated global optimal solution becomes closer to the true global optimum. Therefore, we can obtain a high quality solution by employing MSRM without the added complexity by employing relaxation methods. MSRM obtains different local optima from different initial points. %The best objective value among the local optima is selected as the approximated global optima. The initial points where the optimization starts are determined by the initial value of decision variables. 
To ensure the obtained approximate global minimum is sufficiently close to the true minimum, we employ Latin hypercube sampling to generate many initial points that efficiently cover the entire search space. For each initialization, the proposed optimization problem is solved using IPOPT, a commonly used software that implements an interior point line search filter method to find a suboptimal solution to large nonlinear optimization problems \cite{wachter2009short}.

\section{Numerical simulations} \label{sec:numerical_simulation} 
In this section, SICIN is applied along with two state-of-the-art simulation approaches from Ouyang et al. \cite{ouyang2009methodological} and Fu et al. \cite{fu2016spatial} \footnote{The code and data used for the numerical experiments are available at: \url{https://github.com/YuWVandy/Infrastructure-network-simulation} }. The simulated network samples are compared to two systems of interdependent infrastructure networks. The first consists of real power-water-gas networks in Shelby County, Tennessee, extensively employed in prior research studies \cite{gonzalez2016interdependent}. The second is a simulated system using existing synthetic networks and applied to a region in South China called Xiamen \cite{fang2019adaptive}. The examples are used to benchmark the performance of different simulation methods. Results demonstrate that SICIN outperforms existing methods by producing the most realistic representation of the real system in terms of topological similarity metrics (e.g., average degree, clustering coefficient) and physical quantities (e.g., power load, water flow, and gas pressure). We present a full comparative analysis for the Shelby County system of ICIs, including network visualization, topology, and flow assessment. The Xiamen system is used to demonstrate the generalizability and scalability of SICIN to other systems of different sizes.

\vspace{-5pt}
\subsection{Data and Network Simulation}\label{subsec: data}
Individual networks are first simulated. To ensure that the spatial features of simulated networks are on the same scale as the real networks, we embed the whole system into a specific region. Shelby County in Tennessee spans between ($34.98^{\circ}$N, $35.4^{\circ}$N) in latitude and ($-90.2^{\circ}$W, $-89.6^{\circ}$W) in longitude. The amount of services required for water $z^{{\rm{W}}, t}_i$, gas $z^{{\rm{G}}, t}_i$, and power $z^{{\rm{P}}, t}_i$ at the demand nodes is assumed to be proportional to the population around each demand node. The tract-level population data is obtained from the U.S. Census Bureau. The number of facilities in the simulated gas, power, and water networks is set to be the same as the corresponding networks in Shelby County. The degree sequences of the gas, power, and water networks in Shelby County are used to fit the Poisson distributions in Algorithm~\ref{alg:tripartite_graph}. The second system is located in a coastal city in South China and built using projections of the IEEE 24-bus power system~\cite{grigg1999ieee} and a gas network adapted from the IEEE 9-bus~\cite{ouyang2017mathematical}. The power network contains 10 generation units (supply nodes) and 14 transship (transmission nodes) or load (demand nodes) stations. The gas network contains 3 pumping stations (supply nodes) and 6 deliver stations (demand nodes), both of which are projected on  a $400 \times 400$ km$^2$ area in China. Bus P7 of the power network is taken as a reference node and is located near Xiamen ($24.5^{\circ}$N, $118.0^{\circ}$E) in China~\cite{fang2019adaptive}. %The detailed visualization of the networks is provided in Fig.~1 and~4 of the supplementary material.
\begin{figure}[!htbp]
\centering
\includegraphics[width=\linewidth]{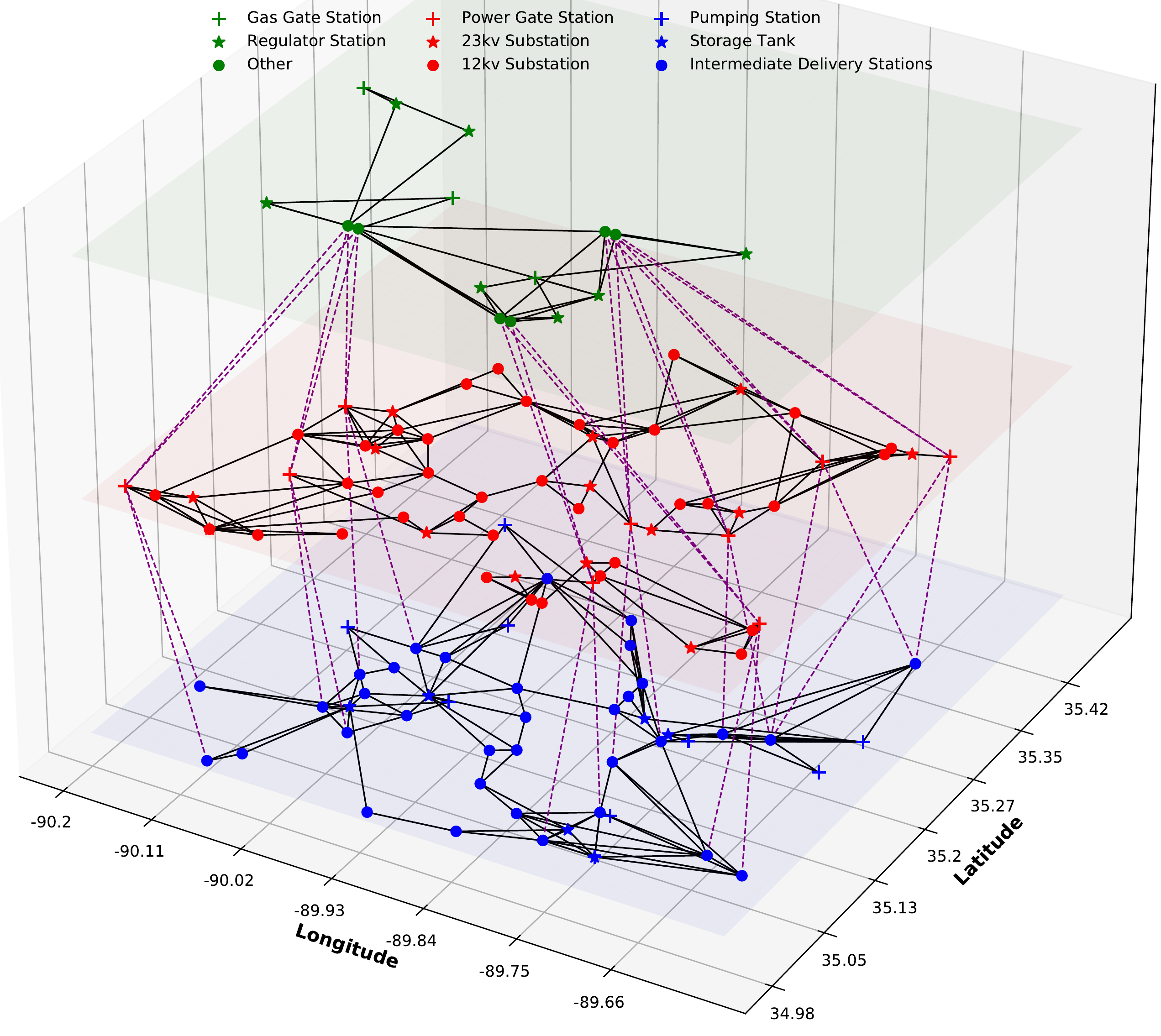}
\caption{The simulated gas-power-water system of Shelby County. The nodes in green, red, and blue represent facilities in gas, power and water networks. Purple lines refer to interdependent links between power, water, and gas nodes.}
\label{Fig: simu_system}
\end{figure}
\\\indent Next, the interdependencies are established between the simulated networks. As described in Section~\ref{sec:ici}, we consider four types of dependencies. The number of nodes on which nodes or links in other networks depend influences the system redundancy and affects the complexity of optimizing the system flow. The system becomes less redundant and more vulnerable to disruptions when the number of such nodes is one, i.e., there is no backup if the dependent facility is disrupted. However, a higher number of such nodes means more interdependent pipelines or power lines, which incurs higher construction fees. Considering the trade-off between the system redundancy and the optimization complexity (construction fees), we set the number of such nodes to two, e.g., in $\rm{P}_{\rm{d}} \to \rm{G}_{\rm{pipe}}$ dependency, every gas pipeline relies on electricity provided by the two nearest 12kV substations.
\\\indent Finally, after we simulate individual networks and their interdependencies, the network flow is initialized with various estimates of parameters for solving the flow optimization obtained from the literature. %listed in Table~\text{I}~in the supplementary material. 
The value of $\delta_1-\delta_5$ in~\eqref{eq:gaspressureflow} is given in~\cite{crane1957flow, gas2004gpsa} and the Weymouth method is used. We assume that the temperature $T_i$ and the adiabatic efficiency $\eta_{ij}$ share the same value across all nodes and arcs. 
\\\indent The simulated network is visualized in Fig.~\ref{Fig: simu_system} which shows one type of interdependent links between power supply nodes and water and gas demand nodes. %Additional types of interdependent links are visualized in Fig.~1(b) of the supplementary material.

\vspace{-5pt}
\subsection{Analysis of Node Locations}\label{subsec: result_annealsimulation}

Tables \ref{table: cost comparison} and \ref{table: cost comparison2} provide a summary of the optimized overall distance for users to access resources in the simulated and real networks. Overall, the convergence time of the MSA algorithm for each of the three networks and each facility type is within one minute, demonstrating the efficiency of MSA for minimizing the overall distance. The iteration upper bound $UB(T_0)$ in MSA is 800, 500, 200, and 100 when the temperature range is [0.1, 0.2], (0.2, 0.3], (0.3, 0.7], and (0.7, 1], respectively. The search space is [0, 5], [0, 3], and [0, 2] when the temperature range is [0.1, 0.3], [0.3, 0.5], and [0.5, 1], respectively. Since the simulated annealing cannot guarantee the global optima, the method by Fu et al. \cite{fu2016spatial} provides a better location assignment of nodes in natural gas networks. The overall distance obtained using Fu et al. \cite{fu2016spatial}, 242.33 km, is less than what SICIN obtains, 359.05 km. And both are less than the distance of the real networks and the method by Ouyang et al. \cite{ouyang2009methodological}. This is the result of applying simulated annealing to optimize the distance in SICIN and the consideration of demand density distribution when determining node locations in Fu et al. \cite{fu2016spatial}. However, neither method considers the geographic constraints that some locations might not be appropriate for building certain types of facilities. For example, in the solved-out optimal distribution plan, a pumping station might be built far from the river or a gate station is planned to be built on sites designed for other purposes. The total distance given by our method will increase and get close to real networks when we consider additional constraints. Similar results are observed for the Xiamen system. Both Fu et al. and SICIN result in a relatively shorter overall distance compared to Ouyang et al. and the real networks. Shorter overall distance indicates higher operating efficiency. As a result, SICIN can also be used to optimize future design of infrastructure systems. 
\begin{table}[htbp!]
\renewcommand{\arraystretch}{1.3}
\caption{Overall distance (km) to access resources for Shelby County}
\label{table: cost comparison}
\centering
% Some packages, such as MDW tools, offer better commands for making tables
% than the plain LaTeX2e tabular which is used here.
\begin{tabular}{lccc}
\hline
\textbf{Method} & \textbf{Water} & \textbf{Power} & \textbf{Gas} \\
\hline
Shelby County & 67.18 &119.07&2059.06\\
SICIN & 37.87 &34.96&359.05\\
Ouyang et al. ~\cite{ouyang2009methodological} & 90.45&80.14&964.35\\
%Centerville & / &/&/\\
Fu et al.~\cite{fu2016spatial} & 47.94 &35.00&242.33\\
\hline
\end{tabular}
\end{table}

\vspace{-5pt}

\begin{table}[htbp!]
\renewcommand{\arraystretch}{1.3}
\caption{Overall distance (km) to access resources for Xiamen}
\label{table: cost comparison2}
\centering
% Some packages, such as MDW tools, offer better commands for making tables
% than the plain LaTeX2e tabular which is used here.
\begin{tabular}{lccc}
\hline
\textbf{Method} & \textbf{Power} & \textbf{Gas} \\
\hline
Xiamen & 263.34&779.23\\
SICIN & 133.83&522.12\\
Ouyang et al. ~\cite{ouyang2009methodological} & 257.13&713.85\\
%Centerville & / &/&/\\
Fu et al.~\cite{fu2016spatial} & 101.03&448.23\\
\hline
\end{tabular}
\end{table}

\vspace{-5pt}
\subsection{Comparison of Network Topology}\label{subsec: result_topology}
We use 6 key topological characteristics to evaluate the performance of SICIN in comparison to the simulation methods by Ouyang et al. \cite{ouyang2009methodological} and Fu et al. \cite{fu2016spatial}. The characteristics include topology and spatial diameter (TD, SD), topology and spatial efficiency (TE, SE), cluster coefficient (CC), and the difference in the adjacency matrices (DA). The diameter, $D$, is the length of the shortest path between the most distanced nodes of a graph. TD measures the topological stretch of the graph while SD measures the geographical stretch of a graph. 
The network efficiency, $E$, assesses how well nodes communicate within networks (i.e., how fast a network mobilizes or delivers the flow of service), and is equal to the inverse of the summation of the shortest topological path lengths between all possible node pairs $i$ and $j$ of the network \cite{watts1998collective}. Since the flow starts from supply nodes and ends at demand nodes in our model, we %do not consider all possible node pairs but 
only consider the pair of supply nodes and demand nodes instead of all possible node pairs. 
The cluster coefficient, CC, measures the degree to which nodes in a graph tend to cluster together. Finally, DA measures the difference between the adjacency matrix (normalized by the network size) of the synthetic networks and that of real networks~\cite{tantardini2019comparing}. Table \ref{table_topology_compare1}-\ref{table_topology_compare2} summarizes topological characteristics of the real and synthetic networks. The results of each simulation method are averaged over 300 realizations and compared to the Shelby County and Xiamen networks. 
\\\indent 
Overall, SICIN generates a realistic synthetic system of ICI networks, leading to the smallest DA. With the exception of the SD, all properties of synthetic networks generated using SICIN are the closest to the real networks of Shelby County. Since the node degree and degree distribution of the network greatly affect the network topology \cite{jing2007effects}, sharing a similar degree sequence results in a similar topology. %the pseudo-tripartite graph to produce the fundamental structure of the generated networks and to add edge connections with the prescribed degree distribution.  
SICIN outperforms all other methods in terms of CC. The value of CC depends significantly on the number of neighborhoods of each node \cite{wang2017comparison} which is determined by the degree sequence. Therefore, a similar degree sequence also ensures a similar CC score. According to the TE, networks generated by SICIN have higher efficiency than those generated by other methods. This is the result of having an upper bound of 2 for the length of most paths in our generated networks. Given the pseudo-tripartite structure, the paths start from the supply node, go through the transmission node, and end at the demand node. In contrast, a lack of constraints in other simulation methods results in many paths with length greater than 2 (i.e., paths going through multiple supply, transmission, and demand facilities), thereby generating networks with TE scores that are substantially lower than the ones of real networks. For SE, our simulated networks have higher efficiency because we use MSA to minimize the overall distance. Finally, the networks generated by other methods have a lower TD than those generated by SICIN. Fu et al. \cite{fu2016spatial} and Ouyang et al. \cite{ouyang2009methodological} employ the preferential attachment assumption that newly added nodes tend to form edges with closer nodes that have a higher degree, and nodes tend to cluster around hub nodes, leading to a smaller topological diameter.

\begin{table}[htbp!]
\scriptsize
\renewcommand{\arraystretch}{1.3}
\caption{Properties of synthetic and Shelby networks}
\label{table_topology_compare1}
\centering
\begin{tabular}{p{4mm}ccccccc}
\hline
\multicolumn{2}{l}{\textbf{Simulation Method}}          & \textbf{CC}     & \textbf{TE}     & \textbf{SE}     & \textbf{TD}     & \textbf{SD}   & \textbf{DA}\\
\hline
\multirow{5}{*}{\textbf{Water}} &\multicolumn{1}{l}{Shelby County}& 0.2221 & 0.1575 & 0.0246 & 6.0000      & 50.7807  & 0\\
& \multicolumn{1}{l}{SICIN}    & \textbf{0.3503} & \textbf{0.1750} & \textbf{0.0259} & \textbf{6.1528} & 64.4045 & \textbf{3.0187}  \\
& \multicolumn{1}{l}{Ouyang et al. ~\cite{ouyang2009methodological}}        & 0.8027 & 0.0663 & 0.0117 & 4.1678 & \textbf{40.9111} & 3.3210\\
& \multicolumn{1}{l}{Fu et al.~\cite{fu2016spatial}}        & 0.7958 & 0.1128 & 0.0060 & 3.6667 & 133.4310 & 4.8599\\
\hline
\multirow{5}{*}{\textbf{Power}} & \multicolumn{1}{l}{Shelby County} & 0.3247 & 0.1485 & 0.0347 & 10.0000     & 71.6696  & 0\\
& \multicolumn{1}{l}{SICIN}    & \textbf{0.3438} & \textbf{0.1378} & \textbf{0.0231} & \textbf{6.1561} & \textbf{59.0578} & \textbf{2.6653} \\
& \multicolumn{1}{l}{Ouyang et al. ~\cite{ouyang2009methodological}}        & 0.7940 & 0.0646 & 0.0114 & 4.4329 & 43.1187 & 3.1364\\
& \multicolumn{1}{l}{Fu et al.~\cite{fu2016spatial}}        & 0.8311 & 0.0937 & 0.0049 & 3.5248 & 128.0882 & 4.7216\\
\hline
\multirow{4}{*}{\textbf{Gas}}   & \multicolumn{1}{l}{Shelby County} & 0.3125 & 0.3611 & 0.0515 & 4.0000      & 66.9558  & 0\\
& \multicolumn{1}{l}{SICIN}    & \textbf{0.3661} & \textbf{0.4850} & \textbf{0.0589} & \textbf{3.1827} & \textbf{56.3581} & \textbf{2.4091} \\
&\multicolumn{1}{l}{Ouyang et al. ~\cite{ouyang2009methodological}}       & 0.8870 & 0.1776 & 0.0182 & 2.0772 & 36.8947 & 2.4261\\
& \multicolumn{1}{l}{Fu et al.~\cite{fu2016spatial}}        & 0.8251 & 0.1346 & 0.0078 & 2.2475 & 90.4244 & 3.1678\\
\hline
\end{tabular}
\end{table}

\vspace{-5pt}

\begin{table}[htbp!]
\scriptsize
\renewcommand{\arraystretch}{1.3}
\caption{Properties of synthetic and Xiamen networks}
\label{table_topology_compare2}
\centering
\begin{tabular}{p{4mm}ccccccc}
\hline
\multicolumn{2}{l}{\textbf{Simulation Method}}          & \textbf{CC}     & \textbf{TE}     & \textbf{SE}     & \textbf{TD}     & \textbf{SD} & \textbf{DA}      \\
\hline
\multirow{5}{*}{\textbf{Power}} & \multicolumn{1}{l}{Xiamen} & 0.2153 & 0.2568 & 0.0031 & 9 & 825.9862 & 0  \\
& \multicolumn{1}{l}{SICIN}    & \textbf{0.3509} & \textbf{0.3352} & \textbf{0.0041} & 3.8857 & 477.1042 & \textbf{2.9129}\\
& \multicolumn{1}{l}{Ouyang et al.}        & 0.7487 & 0.1468 & 0.0022 & 2.9901 & 317.2162 & 3.1999 \\
& \multicolumn{1}{l}{Fu et al.}        & 0.6201 & 0.1543 & 0.0010 & \textbf{4.3023} & \textbf{1056.7987} & 3.8012 \\
\hline
\multirow{5}{*}{\textbf{Gas}} & \multicolumn{1}{l}{Xiamen} & 0.1667 & 0.4861 & 0.0048 & 4 & 403.2344 & 0 \\
& \multicolumn{1}{l}{SICIN}    & \textbf{0.3829} & \textbf{0.6297} & \textbf{0.0060} & 2.3048 & \textbf{401.6185} & \textbf{1.6566}\\
& \multicolumn{1}{l}{Ouyang et al.} & 0.7346 & 0.3682 & 0.0033 & \textbf{2.4158} & 397.3779 & 3.8072 \\
& \multicolumn{1}{l}{Fu et al.} & 0.7639 & 0.2147 & 0.0013 & 2.2591 & 633.6471 & 2.7372\\
\hline
\end{tabular}
\end{table}

\subsection{System Flow Analysis}
%old
% In addition to the comparison of network topology characteristics, we further validate the performance of SICIN by analyzing the synthetic flow to verify the system flow based on initial flow optimization. Three specific gas, power, and water networks are selected from the 300 simulated samples and a system flow optimization is performed. This analysis is done for networks generated based on Shelby County. The optimization is implemented in Julia/JuMP 0.21.2~\cite{dunning2017jump} on a Windows 10 laptop with a 1.9GHz Intel Core i7-8650U CPU and 16 GB RAM. The model is solved by IPOPT which requires the objective function and constraints to be twice differentiable~\cite{wachter2009short}. Therefore, additional constraints are added to impose a non-zero lower bound on the decision variables, $pr_i^t$ and $f_{ij}^t$, since these are in the denominator as a result of the second derivatives of constraints~\eqref{eq:interp2g1}, and \eqref{eq:interWP2}, \eqref{eq:gaspressureflow}.

\begin{figure*}[h]
    \centering % <-- added
\begin{subfigure}{0.25\textwidth}
  \includegraphics[width=\linewidth]{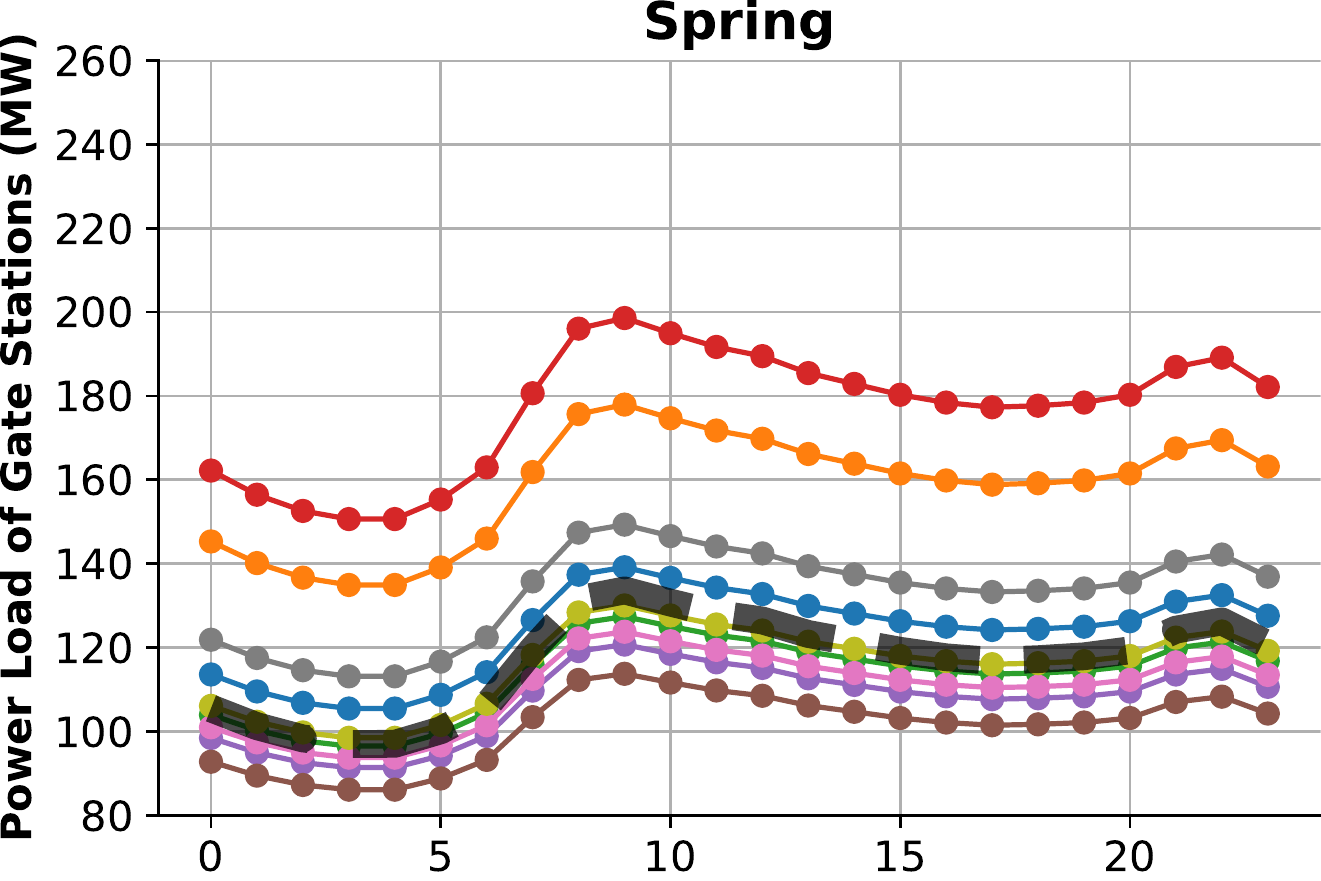}
  \label{fig:1}
\end{subfigure}\hfil % <-- added
\begin{subfigure}{0.25\textwidth}
  \includegraphics[width=\linewidth]{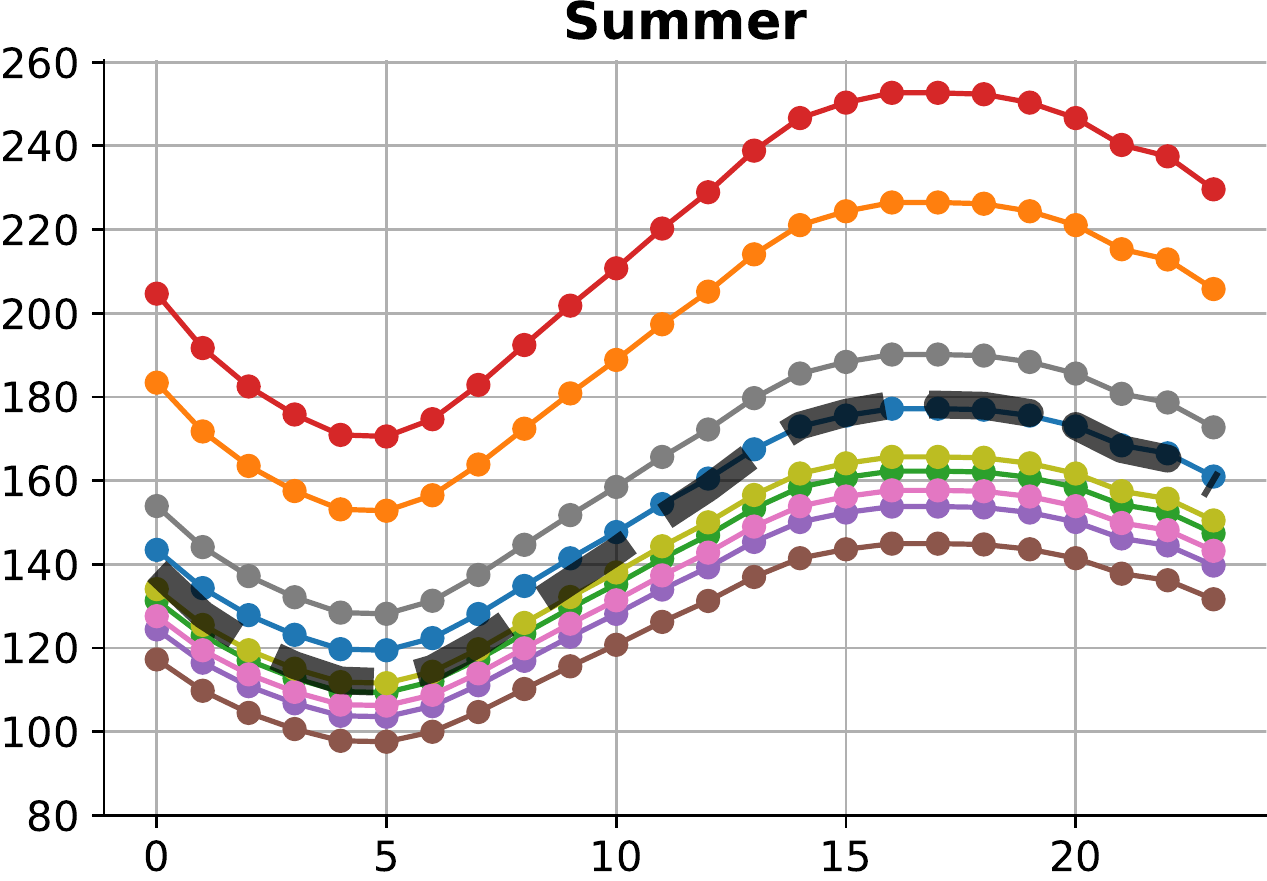}
  \label{fig:2}
\end{subfigure}\hfil % <-- added
\begin{subfigure}{0.25\textwidth}
  \includegraphics[width=\linewidth]{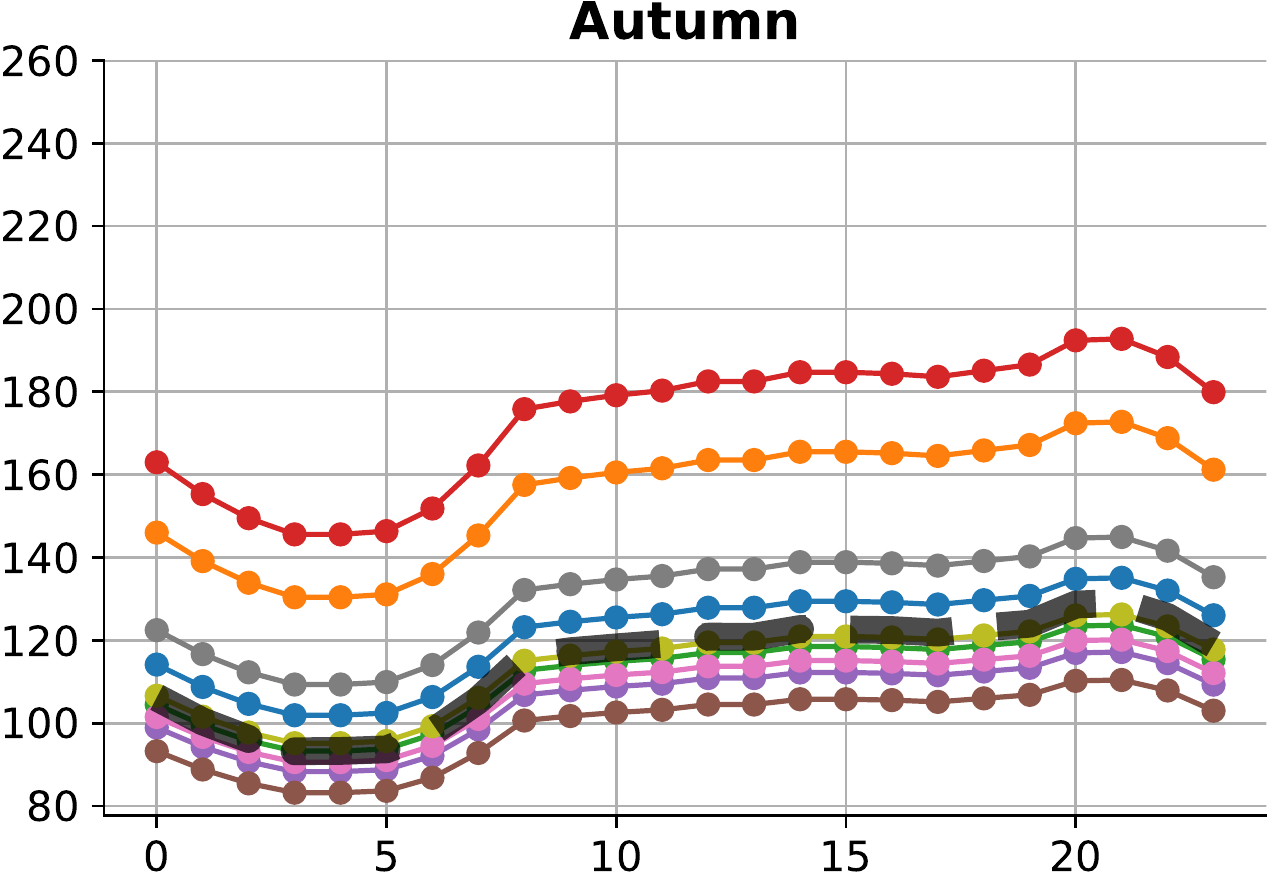}
  \label{fig:3}
\end{subfigure}\hfil
\begin{subfigure}{0.25\textwidth}
  \includegraphics[width=\linewidth]{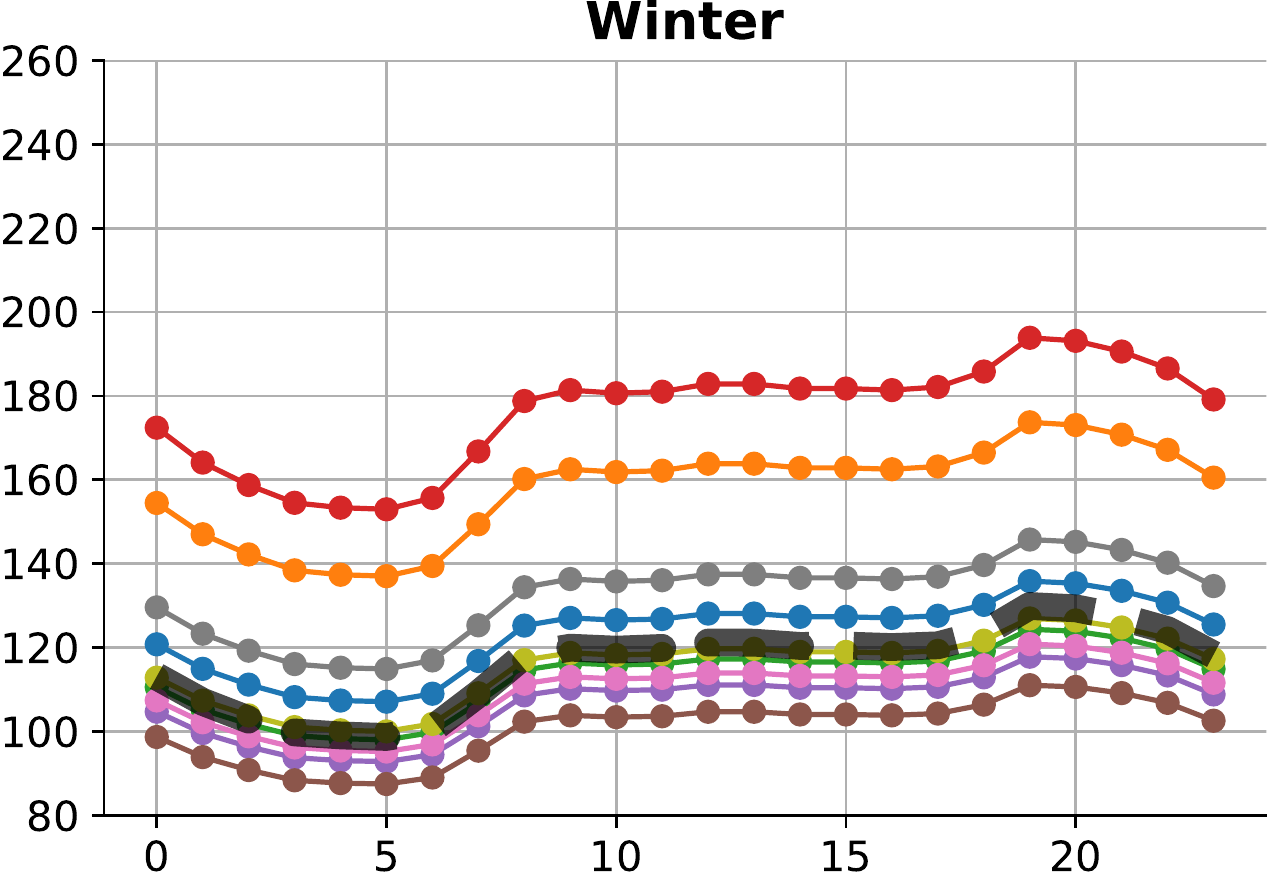}
  \label{fig:4}
\end{subfigure}

\vspace{-5pt}
\begin{subfigure}{0.25\textwidth}
  \includegraphics[width=\linewidth]{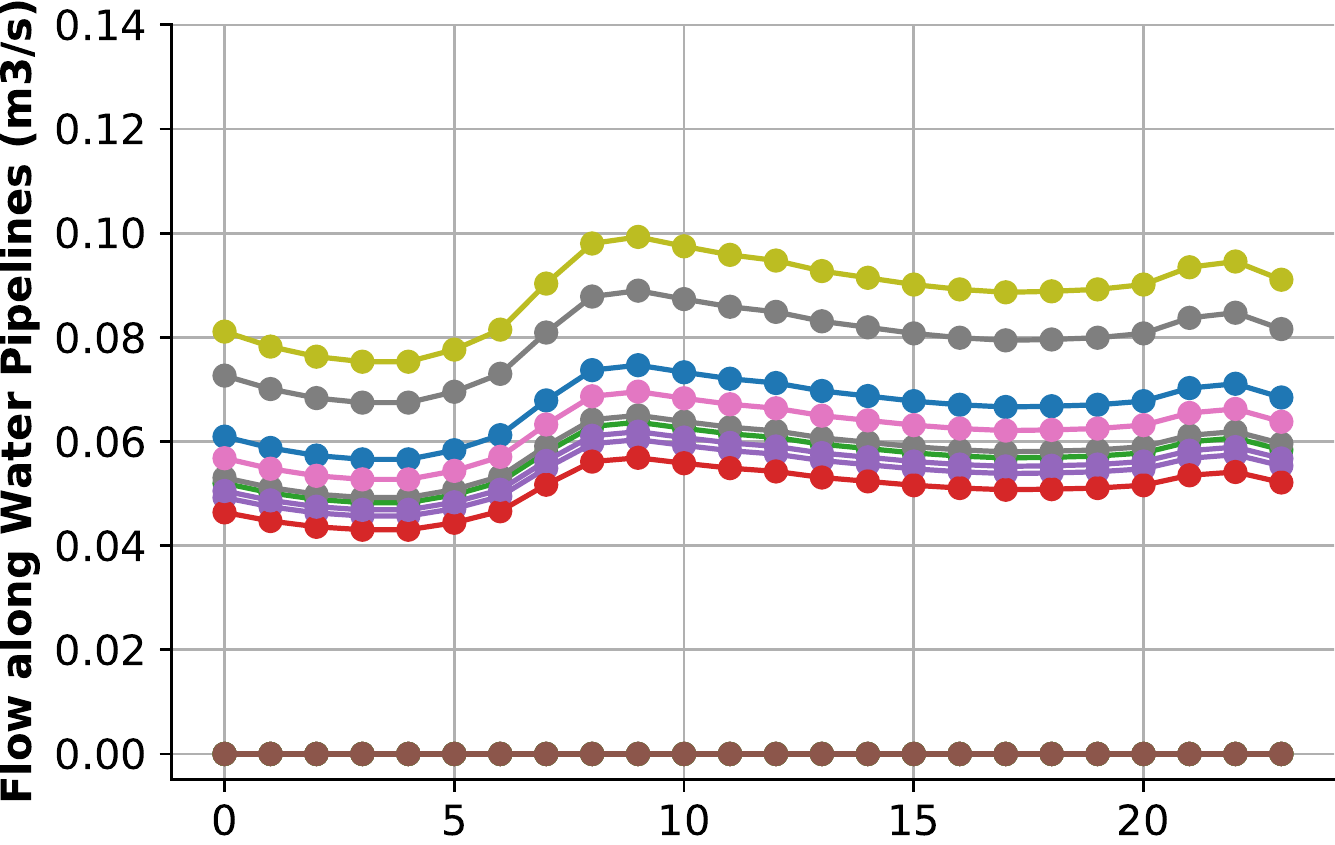}
  \label{fig:5}
\end{subfigure}\hfil % <-- added
\begin{subfigure}{0.25\textwidth}
  \includegraphics[width=\linewidth]{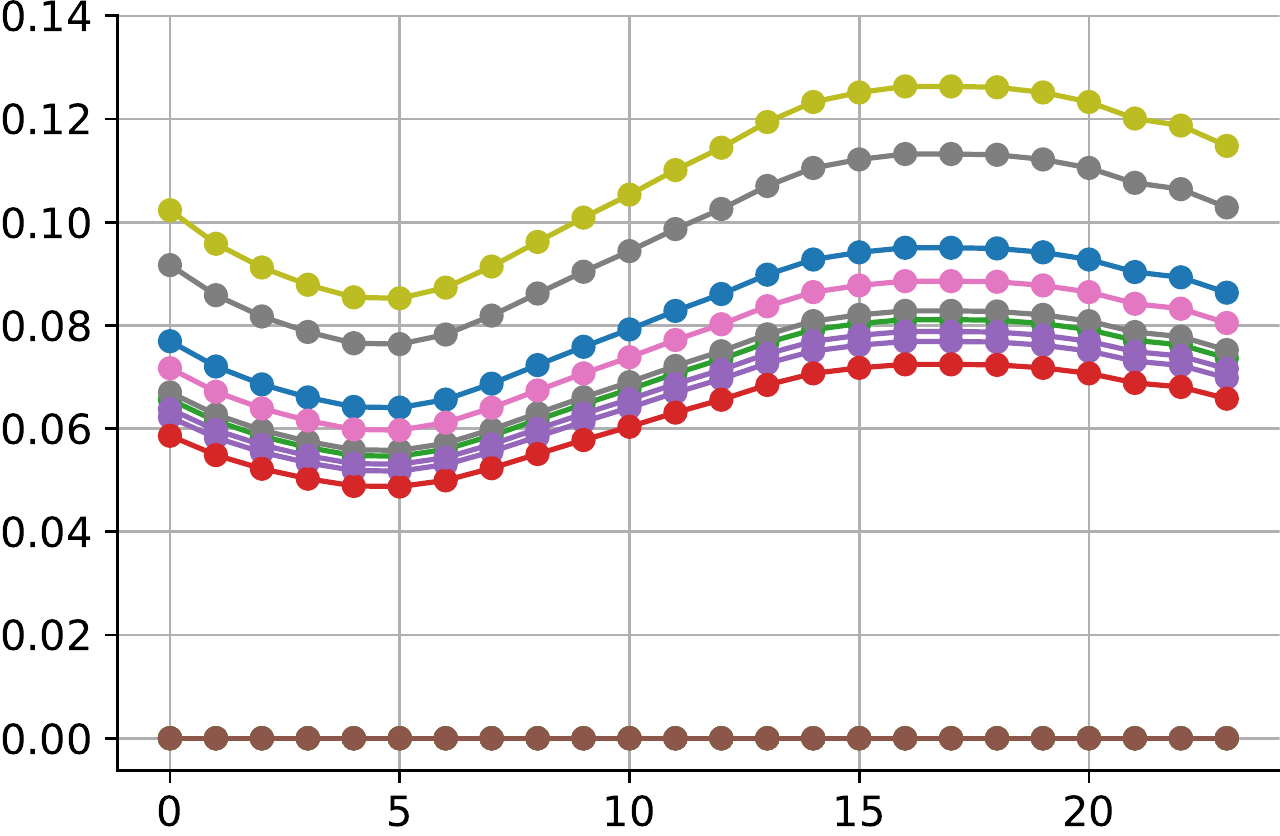}
  \label{fig:6}
\end{subfigure}\hfil % <-- added
\begin{subfigure}{0.25\textwidth}
  \includegraphics[width=\linewidth]{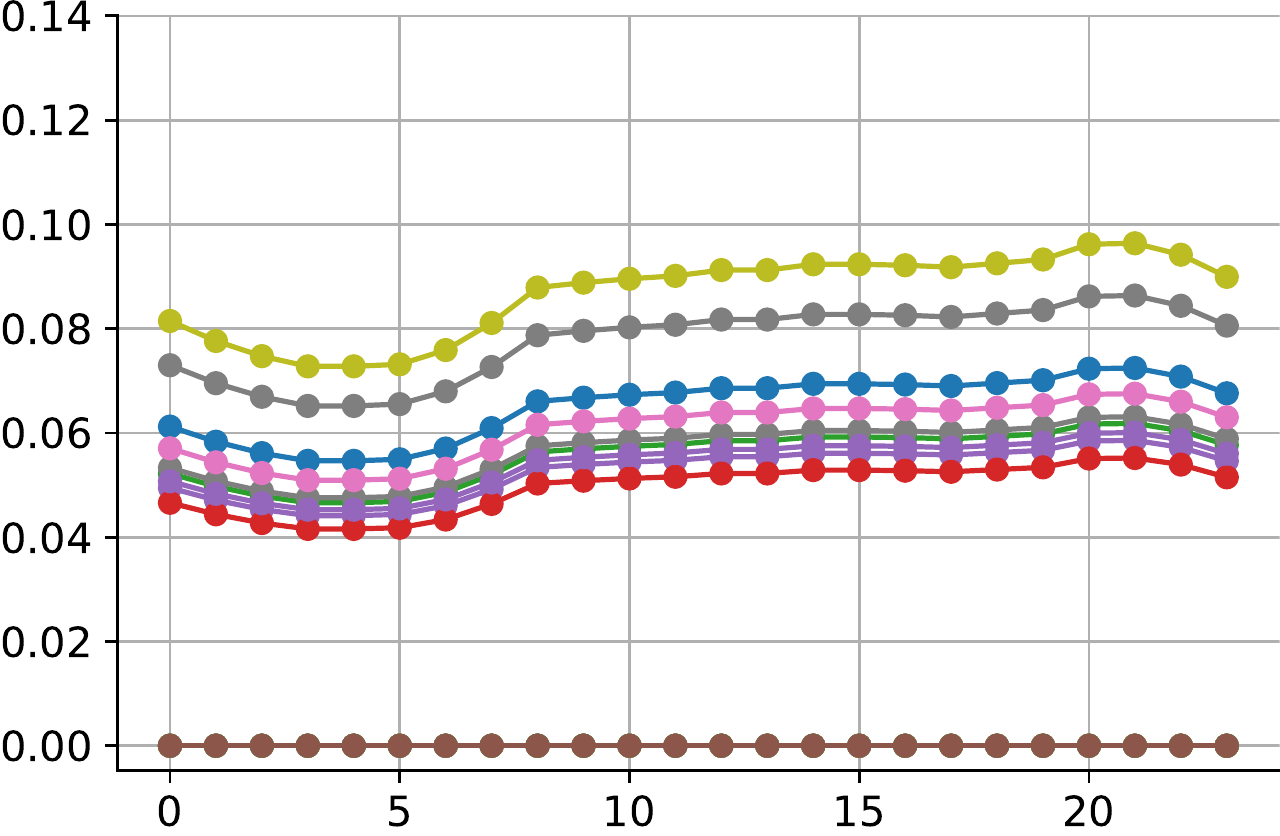}
  \label{fig:7}
\end{subfigure}\hfil
\begin{subfigure}{0.25\textwidth}
  \includegraphics[width=\linewidth]{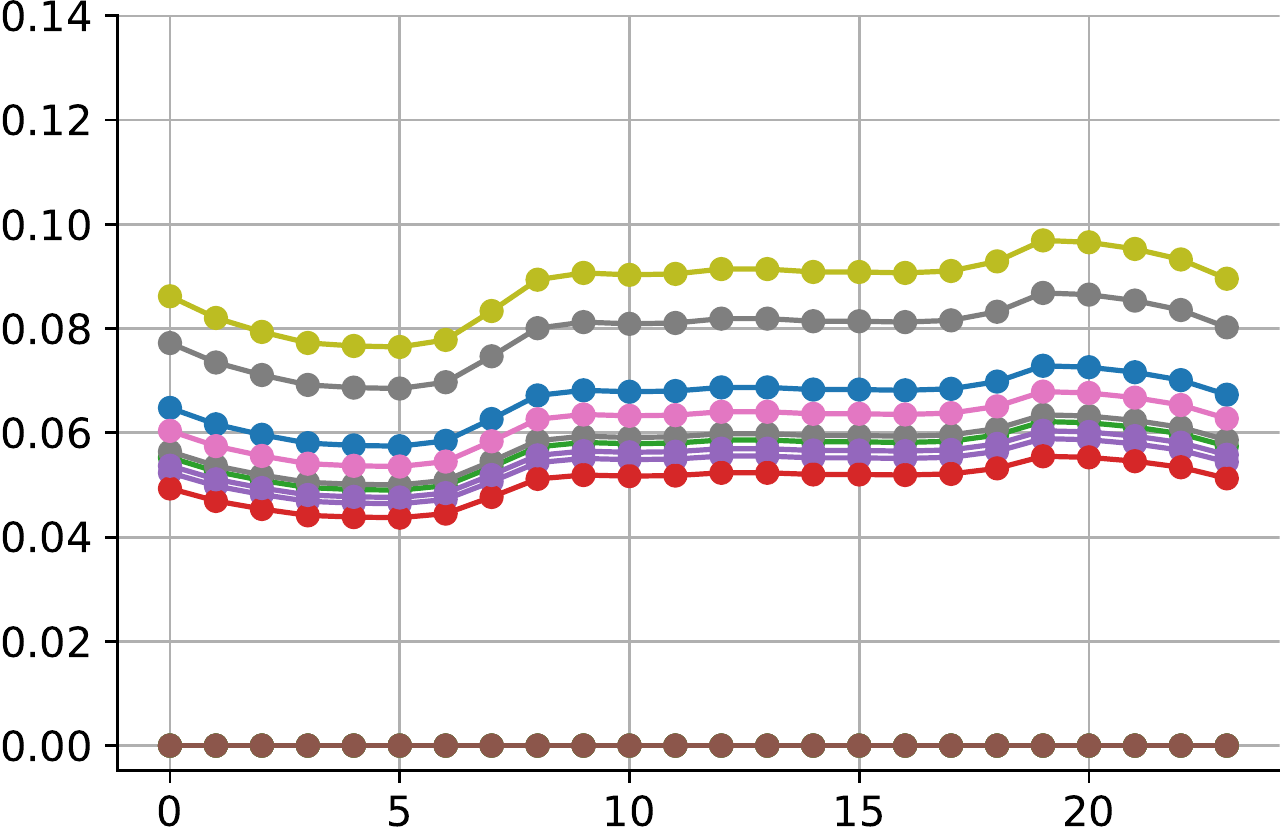}
  \label{fig:8}
\end{subfigure}

\vspace{-5pt}
\begin{subfigure}{0.25\textwidth}
  \includegraphics[width=\linewidth]{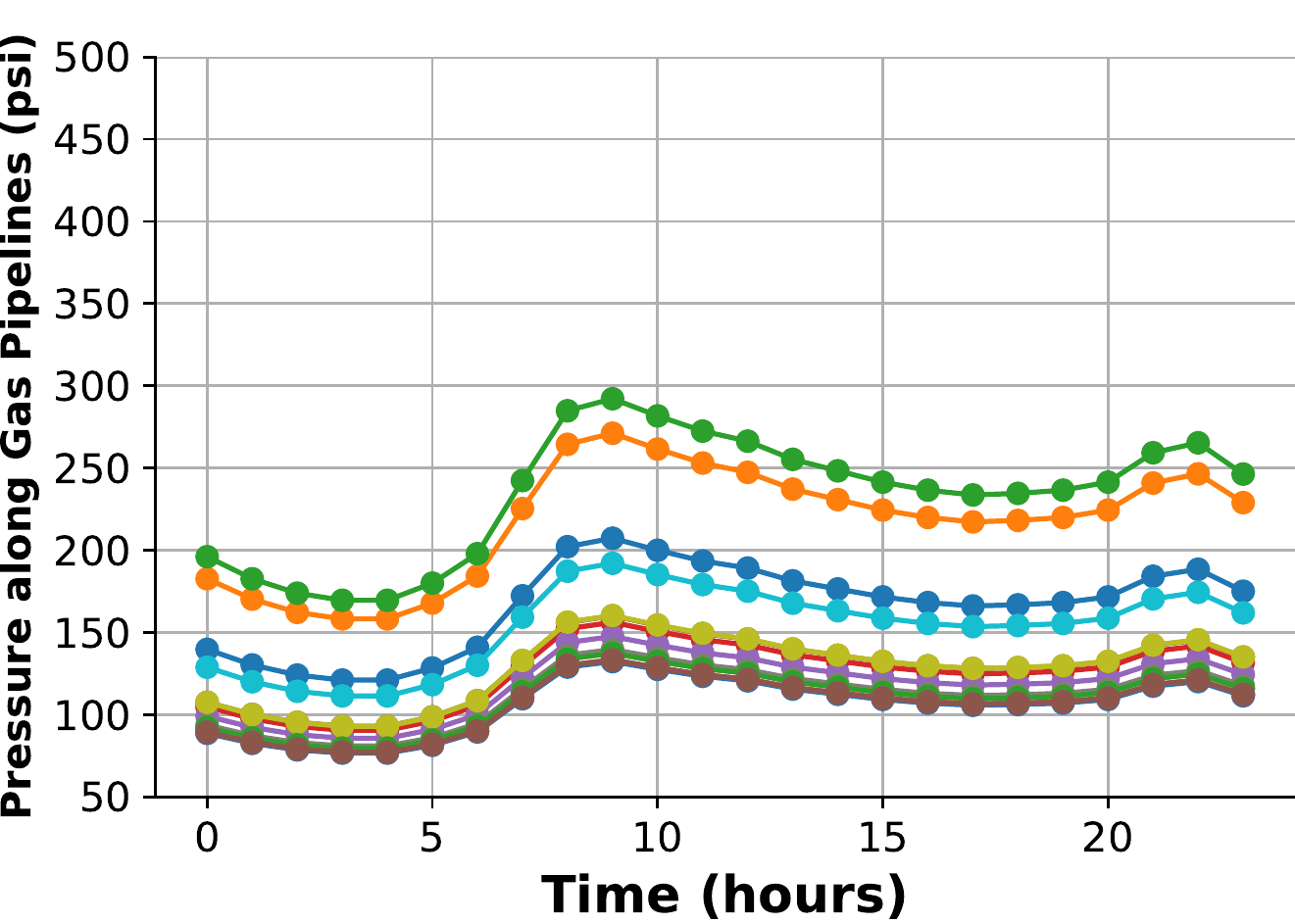}
  \label{fig:9}
\end{subfigure}\hfil % <-- added
\begin{subfigure}{0.25\textwidth}
  \includegraphics[width=\linewidth]{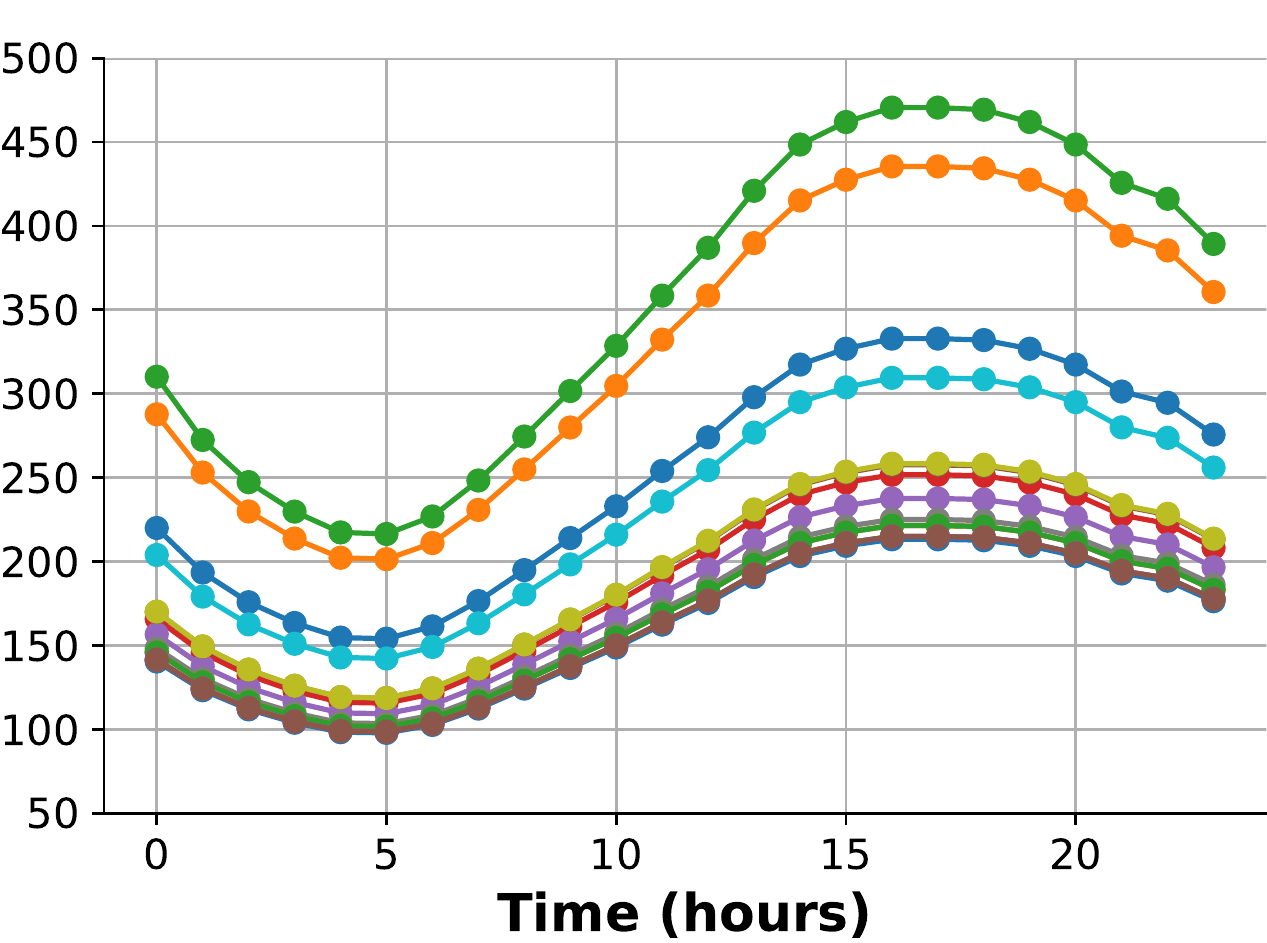}
  \label{fig:10}
\end{subfigure}\hfil % <-- added
\begin{subfigure}{0.25\textwidth}
  \includegraphics[width=\linewidth]{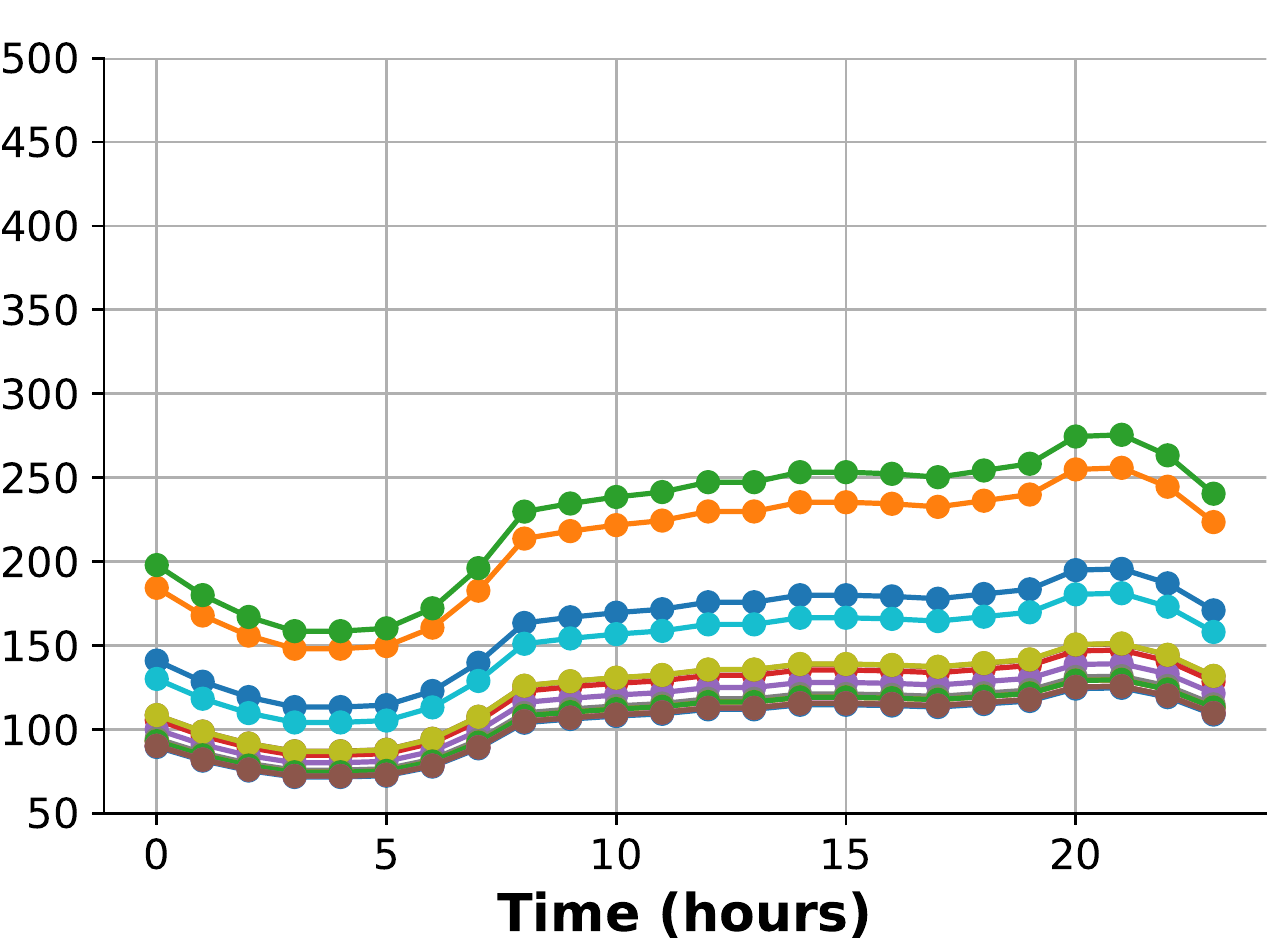}
  \label{fig:11}
\end{subfigure}\hfil
\begin{subfigure}{0.25\textwidth}
  \includegraphics[width=\linewidth]{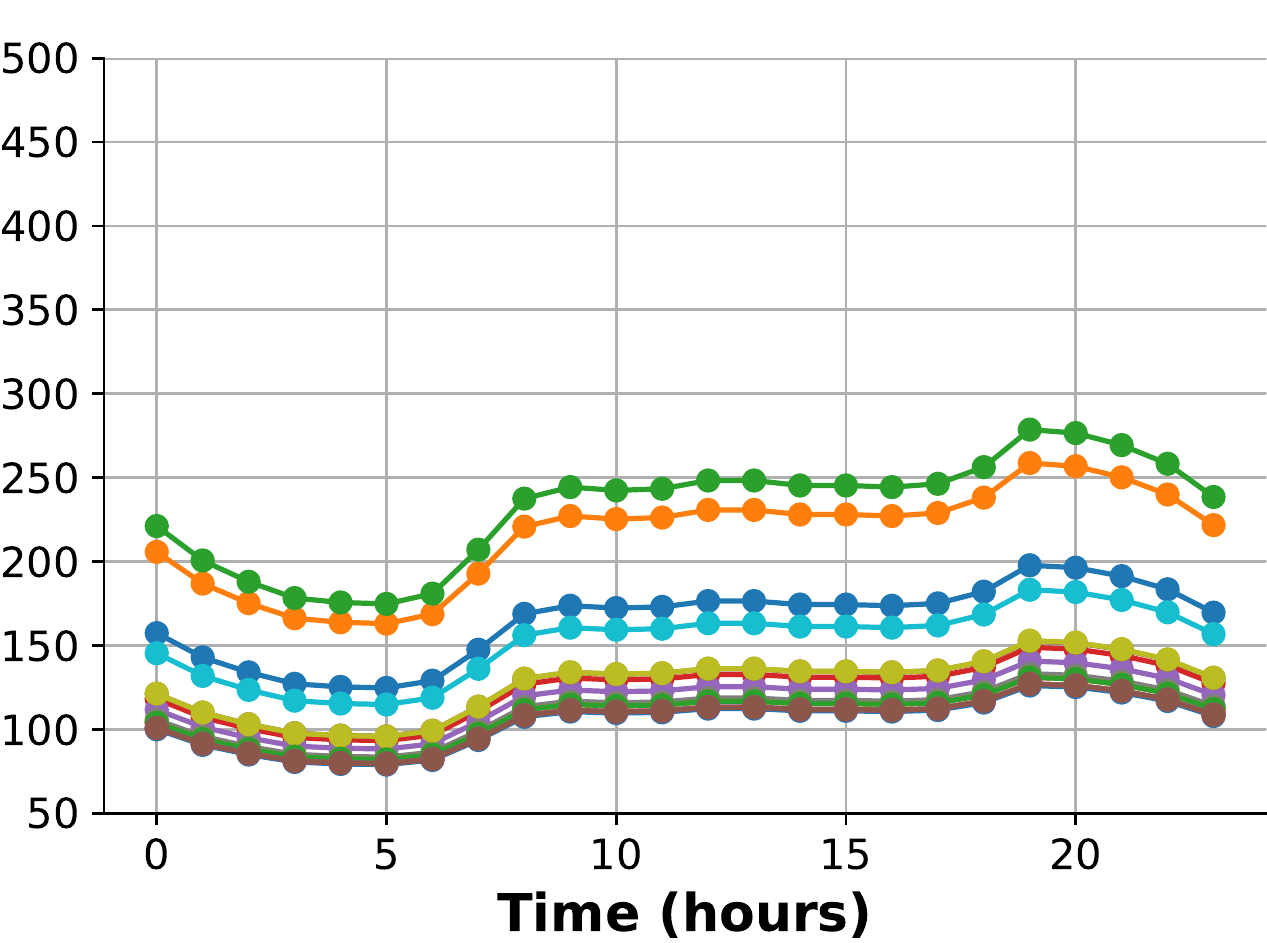}
  \label{fig:12}
\end{subfigure}

\caption{Comparison of optimal solutions from Spring (left) to Winter (right). From top to bottom: Simulated power load for power supply nodes, water flow along interdependent water pipelines from water demand nodes to power supply nodes for cooling purposes, and gas pressure of interdependent gas pipelines from gas demand nodes to power supply nodes. The black dashed lines in figures of power load simulation represent users' electricity demand in Shelby County.}
\label{fig:optimized_result}
\end{figure*}

In addition to the comparison using network topology characteristics, we further validate the performance of SICIN by analyzing the synthetic flow to verify the system flow based on initial flow optimization. Three specific power-gas-water networks are selected from the 300 simulated samples, and a system flow optimization is performed. This analysis is conducted for networks generated based on Shelby County. To ensure that initial points cover the whole feasible space, Latin-hypercube sampling is used considering seven dimensions corresponding to seven decision variables. In each dimension, 1000 points are sampled uniformly from 0 to 1000. For each run, we initialize the decision variable of each dimension using the entry corresponding to that dimension and use IPOPT to get one local optimal solution. This process results in 1000 local optimal solutions, from which we select the minimum as the final optimal solution. The optimization is implemented in Julia/JuMP 0.21.2 on a Windows 10 laptop with a 1.9GHz Intel Core i7-8650U CPU and 16 GB RAM. IPOPT requires the objective function and constraints to be twice differentiable. Therefore, additional constraints are added to impose a nonzero lower bound on the decision variables, $pr_i^t$ and $f_{ij}^t$, since these are in the denominator as a result of the second derivatives of constraints~\eqref{eq:interp2g1}, and \eqref{eq:interWP2}, \eqref{eq:gaspressureflow}.
\begin{equation}
    f_{ij}^t \ge \delta, \forall (i,j) \in ({{\mathcal{A}}^{\rm{W}}}\cup {{{\mathcal{A }}^{{\rm{W} \to \rm{P}}}}})
\end{equation}
\begin{equation}
    pr_i^t \ge \delta, \forall i \in (\mathcal{N}^{\rm{G}} \cup \mathcal{N}^{\rm{P}}_{\rm{s}})
\end{equation}
\indent Four days are selected from each of the four seasons. For each day, the optimization is solved hourly from 00h to 23h using inputs of different power consumption loads to obtain the hourly schedule of each power gate station. We then compare the hourly schedule with users' electricity demand in Tennessee to validate the optimization. For every single optimization, the global optimal is approximated by utilizing the multi-start method and keeping the minimum solution. %To ensure that initial points cover the whole feasible space, Latin-hypercube sampling is used considering seven dimensions corresponding to seven decision variables. In each dimension, one thousand points are sampled ranging from 0 to 1000. 
To obtain the power consumption load, we first calculate the annual average power consumption of each person, $p^{\rm{annual}}$, based on annual electricity consumption per residential customer according to~\eqref{eq_annualp}, where $W^{{\rm{annual}}}$ is the total electricity consumption per residential customer in a year obtained from the U.S. Energy Information Administration (EIA)~\cite{powerconsumptionperson} and $t^{\rm{annual}}$ is the equivalent of one year in seconds. 
\begin{equation}\label{eq_annualp}
    {p^{{\rm{annual}}}} = \frac{{{\rm{W}^{{\rm{annual}}}}}}{{{t^{\rm{annual}}}}}
\end{equation}
Then, the hourly average power consumption in season ``$\rm{sea}$" of each person, $p^{{\rm{sea, hour}}}$, is calculated by multiplying the annual average power consumption, $p^{{\rm{annual}}}$, by the hourly ratio as shown in~\eqref{eq_hourlyp}, where $\frac{{\rm{W}_{{\rm{ave}}}^{{\rm{sea, hour}}}}}{{\rm{W}_{{\rm{ave}}}^{{\rm{annual, hour}}}}}$ is the hourly ratio, and $\rm{W}_{\rm{ave}}^{\rm{sea, hour}}(\rm{W}_{\rm{ave}}^{\rm{annual, hour}})$ represents the seasonal (annual) electricity consumption over an hour. The electricity consumption is calculated by averaging people's total electricity consumption in a season (a year) over the seasonal (annual) time.
\begin{align}\label{eq_hourlyp}
    p^{{\rm{sea, hour}}} &= \frac{{\rm{W}_{{\rm{ave}}}^{{\rm{sea, hour}}}}}{{\rm{W}_{{\rm{ave}}}^{{\rm{annual, hour}}}}}{p^{{\rm{annual}}}}, \\&{\rm{sea}} \in {\rm{\{SP, SU, AU, WIN\} }}, {\rm{hour}} \in [0,23] \nonumber
\end{align}
We select four days from the year 2019 to represent the four seasons, 15-April (SP), 15-July (SU), 15-October (AU), and 15-January (WIN). For each day, we calculate $p^{{\rm{sea, hour}}}$ at different times and solve the optimization to get the four hourly schedules over 24 hours. The schedules are compared with the actual electricity demand value of users in Shelby County. Data on electricity consumption in Tennessee is collected from EIA \cite{powerconsumptionperson}. The data is used to calculate $ {\rm{W}_{{\rm{ave}}}^{{\rm{sea, hour}}}}$ and $\rm{W}_{{\rm{ave}}}^{{\rm{annual, hour}}}$. %The value of $p^{\rm{sea,hour}}$ over time and across seasons is provided in Table~\text{IV} in the supplementary material.
\\\indent Figure~\ref{fig:optimized_result} presents the optimization results, with outcomes from Spring to Winter shown in columns from left to right and each panel representing the power load of gate stations, water flow along water pipelines, and gas pressure in gas pipelines, respectively. For the power load, the black dashed line represents the actual power demand in Shelby County. The colored curves represent the power load at the 9 gate stations. The value at each time point is obtained by dividing the total power demand value by nine which is equal to the number of power supply nodes. Generally, for all four seasons, the power load decreases from 00h to 05h as a result of decreased human activity. The similarity in the pattern of the power load and actual demand from local residents is the result of the power load balancing between power consumption and power generation considering the residents' demand $z_i^{\text{P}, t}$ in constraint \eqref{eq:power cal}. In the summer, the peak occurs around 16h due to increased cooling demand. For the other three seasons, the peak occurs at around 20h, which conforms with typical electricity usage trends. The users' demand curve is within the limits of power load curves from all gate stations. This means that the hourly power load schedule solved by the optimization satisfies users' demand.
\\\indent For water networks, the water flow in all pipelines from Spring to Winter is less than 0.13~m$^3$/s which is below the capacity, 1.13~m$^3$/s, of water pipelines of diameter 0.6~m~\cite{watercapacity}. One pipeline, represented by the brown curve, has a water flow equal to 0 all day and across all four seasons. This is because the dependent number was set to two, Section~\ref{subsec: data}, which means that each power supply node is assigned two water pipelines. In this case, power supply nodes that depend on this water pipeline have satisfied the demand from the other water pipelines. Given that all water pipelines operate below capacity, the cost of transporting water is low, and the optimization has assigned all the flow to the other water pipelines. The pipeline represented by the brown curve, however, can play a critical backup role in the event of a disaster. The trend in water flow over time follows that of the power load due to the interdependency between water and power wherein power supply nodes require water for cooling.
\\\indent Finally, the gas pressure change follows the same trend because electricity is generated by burning natural gas and higher demand for electricity results in higher consumption of natural gas. As the volume of natural gas along pipelines increases, the pressure used for transporting natural gas increases according to~\eqref{eq:gaspressureflow}. Although the natural gas transported through interstate pipelines travels at a pressure ranging from 200 to 1500 psi~\cite{gascapacity}, the solved pressure is for in-state pipelines, which is much lower than interstate pipelines~\cite{gascapacity}. Depending on the type of gas pipelines, the typical pressure could be 100, 125 or 250~psi, which verifies the pressure obtained by our optimization model.

\section{Conclusion and Future Work} \label{sec:conclusion}
% a comprehensive framework was developed to simulate interdependent infrastructure networks. The problem of determining facility location is formulated as an optimal distribution problem and approached by simulated annealing. A pseudo-tripartite graph algorithm is devised to simulate the network topology with prescribed degree distribution. By statistical analysis on network topology, 
The lack of data on interdependent infrastructure systems has limited the ability to understand and model these systems and their interactions to evaluate their vulnerability and resilience. This paper proposes an approach, SICIN, to simulate synthetic ICI networks. SICIN is the first method to consider the simultaneous network flow optimization of multiple infrastructure networks to synthesize interdependent links. 
The contributions of this work include (i) the generated synthetic network that can facilitate future research on ICI systems by providing a benchmark to test and validate models, and (ii) the generalized algorithm that can be used to simulate other infrastructure networks given partial information of real systems. %The synthesized system of interdependent infrastructure networks can facilitate future research on ICI systems by providing a benchmark to test and validate models. %The outcome of this work will enable future research in ICI systems by providing a benchmark to test and validate models using a realistic synthesized system of interdependent infrastructure networks. The contributions of this work include the generated synthetic network and the generalized algorithm which can be used to simulate other infrastructure networks given partial information of real systems. %Our method outperforms current simulation approaches and an analysis of system flow  demonstrates that the constraints introduced in the simulation approach capture realistic trends of water, power, and gas flow.\\

A limitation of SICIN is that it does not consider specific geographic and landscape constraints. The generated locations of infrastructure facilities may not be feasible across different areas. %Additionally, SICIN does not consider other infrastructure, such as road networks, in determining the location of facilities in the infrastructure networks of interest. 
Future work can investigate specific constraints and incorporate other infrastructure such as road networks into the simulation of interdependencies.
Finally, to ensure the scalability of the approach to larger systems, future work can explore other metaheuristics for our problem to improve the convergence speed, such as hybrid metaheuristics~\cite{yuan2017cost}.
% Future work will explore additional types of infrastructure networks such as communication and transportation networks. Further, the current approach cannot guarantee that simulated networks always have a feasible system flow. Therefore, future research can explore network structures such that a feasible flow is guaranteed. Additionally, \tcb{future work can explore other metaheuristics for our problem in the hope to improve the convergence speed, such as hybrid metaheuristics \cite{yuan2017cost}.} %the computation can be improved as using simulated annealing to optimize the distance of obtaining resources is computationally demanding due to the use of brute-force to find nearest neighbors. %The computation can be improved by employing KD-tree.
\appendices
\ifCLASSOPTIONcaptionsoff
\fi
\bibliographystyle{IEEEtran}
\bibliography{main.bib}

\end{document}